\documentclass[twocolumn]{aastex631}

\usepackage{amsmath}
\usepackage{makecell}
\usepackage{gensymb} 

\submitjournal{ApJ}

\begin{document}

\title{Cluster vs Field: Clear Evidence for a Morphology-Density Relation in All Environments at $z\sim1.6$}


\author[0000-0002-6741-078X]{Westley Brown}
\affiliation{Department of Physics and Astronomy, York University, 4700 Keele Street, Toronto, Ontario, Canada, MJ3 1P3}

\correspondingauthor{Westley Brown}
\email{westleyb@yorku.ca}

\author[0000-0002-9330-9108]{Adam Muzzin}
\affiliation{Department of Physics and Astronomy, York University, 4700 Keele Street, Toronto, Ontario, Canada, MJ3 1P3}

\author[0000-0002-7547-3385]{Jasleen Matharu}
\affiliation{Cosmic Dawn Center (DAWN), Denmark}
\affiliation{Niels Bohr Institute, University of Copenhagen, Jagtvej 128, DK-2200 Copenhagen N, Denmark}

\author[0000-0002-6572-7089]{Gillian Wilson}
\affiliation{Department of Physics, University of California Merced, 5200 North Lake Road, Merced, CA 95343, USA}

\author[0000-0002-7356-0629]{Julie Nantais}
\affiliation{Facultad de Ciencias Exactas, Departamento de Ciencias Físicas, Instituto de Astrofísica, Universidad Andrés Bello, Fernández Concha 700, Las Condes, RM 7591538, Chile}

\author[0000-0001-6003-0541]{Ben Forrest}
\affiliation{Department of Physics and Astronomy, University of California,
Davis, One Shields Avenue, Davis, CA 95616, USA}

\author[0000-0003-3921-2177]{Ricardo Demarco}
\affiliation{Institute of Astrophysics, Facultad de Ciencias Exactas, Universidad Andrés Bello, Sede Concepción, Talcahuano, Chile}

\author[0000-0003-1832-4137]{Allison Noble}
\affiliation{School of Earth and Space Exploration, Arizona State University, Tempe, AZ 85287, USA}
\affiliation{Beus Center for Cosmic Foundations, Arizona State University, Tempe, AZ 85287, USA}

\author[0000-0002-0104-9653]{Tracy Webb}
\affiliation{Department of Physics, McGill Space Institute, McGill University, 3600 rue University, Montréal, QC H3A 2T8, Canada}



\begin{abstract}

We explore the relationship between galaxy structure, stellar mass, and local galaxy density in three SpARCS clusters at $z\sim1.6$ and compare with field galaxies from the 3D-HST survey.
Our cluster and field data include: 1) unprecedented multiband photometry, allowing for accurate stellar mass estimates; 2) extensive slit and grism spectroscopy targeting both star-forming and quiescent galaxies, allowing for high-accuracy local density measurements; and 3) deep imaging in F160W, allowing for accurate rest-frame optical morphologies.
Using Sérsic index measured in rest-frame R-band, we classify galaxies as disk-like, bulge-like, and intermediate.
Our sample includes 111 cluster galaxies and 458 field galaxies with reliable Sérsic measurements.
We find that a morphology-density relation is already in-place in both cluster and field galaxies at $z\sim1.6$, such that as local density increases, the fraction of bulge-like galaxies increases and disk-like galaxies decreases.
Both samples show similar positive trends between median Sérsic index and local density.
Additionally, we find a general positive relationship between Sérsic index and stellar mass.
The majority of galaxies remain disk-like until reaching stellar masses above $10^{10.25} M_{\odot}$ in the cluster or $10^{10.8} M_\odot$ in the field, however, we cannot conclude whether the differences in stellar mass trends are significant.
Overall, our results show clear morphology-density and morphology-mass relations in place at $z\sim1.6$ and oppose the idea that cluster-specific processes are solely responsible the morphology-density relation.
Our data further suggest that the morphology-density relation may be independent of global environment at this epoch.

\end{abstract}




\section{Introduction}
\label{sec:intro}


Morphology is one of the most striking features we can observe about the galaxies around us. For decades, astronomers have classified morphology using the Hubble sequence, separating galaxies into early-types such as elliptical and lenticular galaxies, and late-types such as spiral galaxies.
While the Hubble sequence is based purely on visual appearance, morphology has been found to correlate with a range of galaxy properties, including stellar mass \citep{Kauffmann2003}, size \citep{vanderWel2014}, color \citep{deVaucouleurs1961, Strateva2001, Cassata2007}, star formation rates \citep{Brinchmann2004, Bell2012}, and star formation histories \citep{Strateva2001}.
Morphology is also correlated with kinematic properties of galaxies such as rotational velocity \citep{Freeman1970}, angular momentum \citep{Harrison2017}, and velocity dispersion \citep{Djorgovski1987}.

The relationship between morphology and local or global environments can also reveal clues about galaxy evolution.
By studying visual morphology in 55 galaxy clusters in the nearby universe, \citet{Dressler1980} found the first evidence of the morphology-density relation: as local galaxy density increases, there is an increase in the fractions of both elliptical and lenticular galaxies and a decrease in the fraction of spiral galaxies.
This general result has been discovered across many studies and clusters (e.g. \citealp{vanderWel2008} and \citealp{Vulcani2023} at $z<0.1$; \citealp{Dressler1997}, \citealp{Fasano2000}, and \citealp{Goto2003} up to $z\sim0.5$; \citealp{Postman2005} and \citealp{Smith2005} up to $z\sim1$) with many finding that the relation appears to flatten as redshift increases.
The morphology-density relation has become a fundamental part of our understanding of galaxy clusters and environmental evolution, although questions still remain about how far into the cosmic past the relation can be found.

Cosmic noon ($1<z<3$) is an important epoch in galaxy evolution, representing the peak of star formation in the universe, and the period where early-universe protoclusters begin to transform into present-day galaxy clusters.
Studies such as \citet{Tasca2009} and \citet{Euclid2025} have made progress using self-consistent methods to measure the redshift evolution of the morphology-density relation, but their work remains limited to $0.2<z<1$ and $0.25<z<1$, respectively.
Few studies have explored the relationship between morphology and local galaxy density at $z>1$ (e.g. \citealp{Sazonova2020,Mei2023,Strazzullo2023}), and those that have suffer drawbacks such as limited photometric filter coverage, inconsistent rest-frame wavelengths, and uncertain stellar masses.
Photometry limitations and lack of galaxy spectroscopy also means that many of these studies lack accurate galaxy redshift estimates, which reduces the ability to determine cluster membership and calculate accurate local galaxy densities.
Visual classification of galaxy morphology also becomes more difficult at high redshift due to lower image resolution and surface brightness dimming, leading to greater uncertainty in classifications and difficulty distinguishing elliptical and S0 galaxies (e.g. \citealp{Cerulo2017}).

Visual classification has been standard for many years, even adapting to large volumes of data with large-scale citizen science projects such as Galaxy Zoo (e.g. \citealp{Bamford2009}).
However, automated measurements of galaxy structure have gained popularity as quantitative alternatives to manual classification.
Parametric techniques such as Sérsic profile fitting \citep{Sersic1968} and bulge-disk decomposition \citep{Simard2011}, as well as non-parametric techniques such as concentration, asymmetry, smoothness/clumpiness (CAS; \citealp{Conselice2003}), and Gini-M$_{20}$ statistics \citep{Abraham2003, Lotz2004}, all offer benefits over visual classification methods in terms of efficiency, classification bias, and error estimation.
It is common to use these measurements as proxies for visual morphological classification by assigning traditional classifications to specified values or locations in parameter space.
However, quantitative techniques also offer an opportunity to study the full distribution of galaxy structure without relying on discrete categorizations (e.g. \citealp{Sazonova2020}).
At high-$z$, these techniques could be used to help us measure changes in galaxy structure independent of the low-redshift Hubble sequence.

Among other remaining gaps in our knowledge are the role of local environment (e.g. local galaxy density) compared to global environment (e.g. clusters, groups, fields).
As a phenomenon, the morphology-density relation has generally been associated with galaxy clusters, due to the degeneracy between local density and clustercentric radius in relaxed clusters \citep{Whitmore1991,Whitmore1993,Fasano2015,Vulcani2023}.
Despite this broad association, results from studies sampling galaxies across a wider range of environments support the idea that the morphology-density relation may extend beyond the cluster environment \citep{Postman1984,Goto2003,Bamford2009,Tasca2009,Euclid2025}.
Using galaxy samples from the Sloan Digital Sky Survey (SDSS), \citet{Goto2003} suggest that the morphology-density relation may be driven by a two different mechanisms, which act separately on local and global scales.
On the star formation end, \citet{Perez-Millan2023} suggest that local effects may be more important than large-scale environment in determining the stellar properties of galaxies at fixed morphology and stellar mass.
Investigating the impact of different global environments may provide a key to understanding the physical origins of the morphology-density relation.

Galaxy stellar mass is another key factor to consider. Stellar mass is strongly correlated with a wide range of galaxy properties, including morphology \citep{vanderWel2008,Vulcani2011,Cerulo2017,Euclid2025}.
Previous studies have also found positive correlations between stellar mass and local galaxy density \citep{Bamford2009,Muzzin2012}, leading to questions of whether the morphology-density relation may be driven by a more fundamental relation with galaxy stellar mass.
Many studies have sought to assess the separate impacts of mass and environment on galaxy quenching \citep{Peng2010b, Muzzin2012,Kawinwanichakij2017}, but similar analysis has been largely neglected when it comes to galaxy morphology.
At $z\lesssim0.1$, \citet{Calvi2012} find that galaxies residing in clusters follow a different morphology-mass relation than isolated or group galaxies.
To answer to this question requires further analysis of the how the morphology-mass relation depends on global environment.

In this study, we investigate the relationships between galaxy morphology, local density, and stellar mass at $z\sim1.6$, and explore how they are shaped by global environment.
We perform this study using samples with an unprecedented amount of multiband photometric data and slitless grism spectroscopy at $z\sim1.6$, and utilizing Sérsic fitting to obtain quantitative measurements of galaxy structure.
By examining cluster and field samples separately, we not only explore whether the morphology-density relation is in-place by the end of cosmic noon, but gain a greater understanding of the mechanisms that drive it.

We adopt a $\Lambda$CDM cosmology with $H_0 = 70$ km s$^{-1}$ Mpc$^{-1}$, $\Omega_m = 0.3$, and $\Omega_{\Lambda} = 0.7$.
In Section \ref{sec:data}, we describe our data and samples, including clusters, fields, grism data reduction, and galaxy selection. We summarize our structural analysis in Section \ref{sec:analysis}, including Sérsic fitting and local density calculations. In Section \ref{sec:results}, we present our results for the morphology-density and morphology-mass relations in both samples at $z\sim1.6$. We discuss potential physical interpretations of these results in Section \ref{sec:discussion}, and summarize our conclusions in Section \ref{sec:conclusions}.


\section{Data}
\label{sec:data}


\subsection{SpARCS Clusters}
\label{subsec:clusters}

The galaxy clusters used in this study come from the \textit{Spitzer} Adaptation of the Red-sequence Cluster Survey (SpARCS, \citealp{Muzzin2009,Wilson2009}).
With a goal of detecting a large homogeneous sample of $z \sim 1$ galaxy clusters, SpARCS utilized existing observations of the \textit{Spitzer} Wide-Area Infrared Extragalactic Survey (SWIRE) Legacy fields to find cluster candidates.
In particular, our $z\sim1.6$ clusters were detected using the stellar bump sequence outlined in \citet{Muzzin2013}.
This method uses a combination of 3.6 $\mu$m--4.5 $\mu$m and a $z'$--3.6 $\mu$m color cut to detect overdensities at $1.4<z<1.7$ with reasonably tight photometric redshift estimates.

We use three spectroscopically-confirmed SpARCS clusters for this study: J033056-284300 (J0330) at a cluster redshift $z_\textrm{cl}=1.626$, J022426-032330 (J0224) at $z_\textrm{cl}=1.633$, and J022546-035517 (J0225) at $z_\textrm{cl}=1.594$.
Clusters J0330 and J0224 were first presented in \citet{Lidman2012}, while spectroscopic confirmation of J0224 was presented in \citet{Muzzin2013}. J0225 was first presented in \citet{Nantais2016}.
These clusters are ideal for our study as they have: 1) an extensive set of spectroscopic redshifts, including MOSFIRE observations of H$\alpha$ emitters; 2) follow-up \textit{HST}/WFC3 G102 slitless grism observations targeting the Balmer/4000\AA\ break to determine spectroscopic redshifts of quiescent galaxies; and 3) extensive multiband photometry to complete the lower-mass end of the galaxy sample with well-constrained photometric redshifts.
The clusters also have deep imaging in the F160W filter from the See Change program \citep{Hayden2021}, which probes the rest-frame R-band and gives accurate stellar morphologies.
Additional details on the cluster data are provided in the following sections.


\subsubsection{Cluster Photometry, Spectroscopy, and HST WFC3 Imaging}
\label{subsec:cluster_data}

Table \ref{tab:clusters} provides a summary of the observational data for each cluster.
All three clusters have photometric coverage in more than 19 total filters, from optical/near-IR bands (\textit{ugrizYKs} and F160W) up to infrared/far-IR (3.6/4.5/5.8/8.0/12/24/110/160/250/350/500 $\mu$m).
J0330 and J0224 also have photometry in \textit{J}, F814W, F105W, and F140W.
Additional observations of each cluster from ALMA in CO (2--1) are presented in \citet{Noble2017}, while observations from HST/ACS in F475W and F625W are presented in \citet{Cramer2023}, although they are not used in this study.

\begin{deluxetable*}{lccc cccc}
    \tabletypesize{\footnotesize}
    \tablecolumns{8}
    \tablecaption{Summary of available data for the SpARCS clusters used in this dataset. \label{tab:clusters}}
    \tablehead{
        \colhead{Cluster ID} & \colhead{R.A.} & \colhead{Decl.} & \colhead{$z_\textrm{cl}$} & \colhead{\makecell{Optical/Near-IR \\ Photometry}} & \colhead{\makecell{IR/Far-IR \\ Photometry}} & \colhead{Spectroscopy} & \colhead{Grism}
    }
    \startdata
        SpARCS J0330 & $03^\textrm{h} 30^\textrm{m} 55.9^\textrm{s}$ & $-28\degree 42' 59.5''$ & 1.626 & \makecell{\textit{ugrizYJKs} F814W, \\ F105W, F140W, F160W} & \makecell{[3.6] [4.5] [5.8] [8.0] [12] [24] \\ \ [110] [160] [250] [350] [500]} & \makecell{MOSFIRE,\\FORS2, OzDES} & G102\\
        \hline
        SpARCS J0225 & $02^\textrm{h} 25^\textrm{m} 45.6^\textrm{s}$ & $-03\degree 55' 17.1''$ & 1.594 & \textit{ugrizYKs} F160W & \makecell{[3.6] [4.5] [5.8] [8.0] [12] [24] \\ \ [110] [160] [250] [350] [500]} & \makecell{MOSFIRE,\\FORS2, OzDES} & G102\\
        \hline
        SpARCS J0224 & $02^\textrm{h} 24^\textrm{m} 26.3^\textrm{s}$ & $-03\degree 23' 30.8''$ & 1.633 & \makecell{\textit{ugrizYJKs} F814W, \\ F105W, F140W, F160W} & \makecell{[3.6] [4.5] [5.8] [8.0] [12] [24] \\ \ [110] [160] [250] [350] [500]} & \makecell{MOSFIRE,\\FORS2, OzDES} & G102\\
    \enddata
    \tablecomments{Coordinates and redshifts are based on the BCG of the cluster. While not used in this dataset, additional data for these clusters include observations from ALMA in CO (2--1) \citep{Noble2017}, and observations from \textit{HST}/ACS in F475W and F625W \citep{Cramer2023}.
    }
\end{deluxetable*}

Ground-based spectroscopy of each cluster comes from Keck/MOSFIRE, VLT/FORS2, and OzDES.
Targets of the MOSFIRE spectroscopy are primarily H$\alpha$ emitters with color-color cuts consistent with the cluster redshift.
\citet{Nantais2016} identify 38 spectroscopically-confirmed cluster members in J0330, 8 in J0225, and 45 in J0224.
An additional 6 cluster members are spectroscopically confirmed in \citet{Nantais2020} through modified reductions of the MOSFIRE observations.
We direct the reader to \citet{Nantais2020} for details on the MOSFIRE masks and data reduction.
Details on the data reduction of the WFC3 G102 slitless spectroscopy and confirmation of grism cluster members are given in Section \ref{subsec:grism}.

Photometric catalogs for each cluster include photometric redshift estimates and rest-frame \textit{UVJ} colors from EAZY \citep{Brammer2008}, as well as galaxy stellar mass estimates from FAST \citep{Kriek2009}. We refer the reader to \citet{Nantais2016} and \citet{Nantais2020} for a full discussion on the creation of the photometric catalogs.

For our primary analysis and structural measurements, we utilize drizzled images from WFC3/IR F160W, which have been reduced to a plate scale of 0.06\arcsec/px. At $z\sim1.6$, this covers a rest-frame wavelength of roughly $550-650$ nm and is equivalent to the rest-frame optical R-band.
F160W observations for all three clusters were obtained from GO-13306 (PI Wilson), while additional F160W observations for J0330 and J0224 were obtained as part of the See Change program \citep{Hayden2021}. This results in total F160W exposure times of 5019 seconds and 6116 seconds for J0330 and J0224, respectively, and 2424 seconds for J0225.
The drizzled F160W images cover areas of roughly 2\arcmin$\times$2\arcmin\ targeted at the center of each cluster, which is smaller than the coverage of most ground-based observations of the clusters.
It is important to note that some objects in the catalogs, including spectroscopically-confirmed cluster members, lie beyond the bounds of the F160W imaging and therefore are not included in our analysis.


\subsubsection{WFC3 G102 Grism Data Reduction}
\label{subsec:grism}

All three clusters in our sample have data from the WFC3 G102 grism to a depth of 4 orbits in two orients (GO-13306; PI Wilson).
Slitless grism spectroscopy allows one to obtain low-resolution spectra of all objects in a field of view at once.
The G102 grism covers an observed wavelength range of $800-1150$ nm, with a resolving power of 210 at 1000 nm and a dispersion of $2.45$ nm/pixel.
At $z \sim 1.6$, the 4000\AA\ break and [O\textsc{ii}] emission line doublet (rest-frame $\lambda \lambda3726-3729$\AA) shift to observed wavelengths of 1050 nm and approximately 970 nm, respectively.
The 4000\AA\ break is dominant in galaxies with older stellar populations, while [O\textsc{ii}] emission is caused by younger stellar populations. In combination with the extensive MOSFIRE spectroscopy targeting H$_\alpha$, the grism observations were designed to provide a more complete sample of both star-forming and quiescent cluster galaxies.

We reduce the G102 data of our clusters using \texttt{grizli}: grism redshift \& line analysis software for space-based slitless spectroscopy \citep{Grizli}. The \texttt{grizli} pipeline is designed for end-to-end data reduction and processing of grism exposures from both \textit{HST}/WFC3 and \textit{JWST}/NIRISS.
\texttt{Grizli} uses traces of the objects to model the object continuum and background contamination (such as from neighboring objects or spurious detections) separately.
The pipeline creates an object catalog from the corresponding direct detection image, which is then used to extract the 2D spectra of individual objects from the full grism field of view.
\texttt{Grizli} can then model and fit the spectrum of each object, producing a 1D spectrum and fitting a redshift to the observed spectral features.

Due to the small wavelength range of G102, it is difficult to accurately constrain the redshift or galaxy SED from the grism spectrum alone. We assume any redshift estimates based on G102 alone to be uncertain and unreliable, and therefore process our data using the available photometry for each object. The wide wavelength coverage of our photometry improves the accuracy in modeling galaxy SEDs and correctly identifying spectral features.

Before inspecting the grism results, we begin by matching the \texttt{grizli} catalog to objects in our cluster catalogs.
We eliminate from our grism sample all objects for which \texttt{grizli} failed to produce a fit with photometric data points.
We then collect all redshift estimates of each object, including photometric, grism, and spectroscopic (if available).
If at least one of the three redshift estimates lies near the range of the clusters (roughly $1.4<z<1.8$), we keep the object in the grism sample for inspection.

We assess the reliability of the fit of each remaining object by visually inspecting the galaxy SEDs, 1D grism specta, and 2D grism spectra generated by \texttt{grizli}.
We note that even if a grism spectrum is faint or partially contaminated, \texttt{grizli} may successfully fit an SED if the object has good photometry.
We flag the grism redshift as unreliable for objects whose spectrum appears poorly resolved or does not include visible features such as a 4000\AA\ break or [O\textsc{ii}] emission line, as well as for objects whose spectral trace appears badly contaminated.
After inspection, the reliable grism redshifts are incorporated into our cluster catalogs.

\begin{figure*}[tb]
    \centering
    \includegraphics[width=1\linewidth]{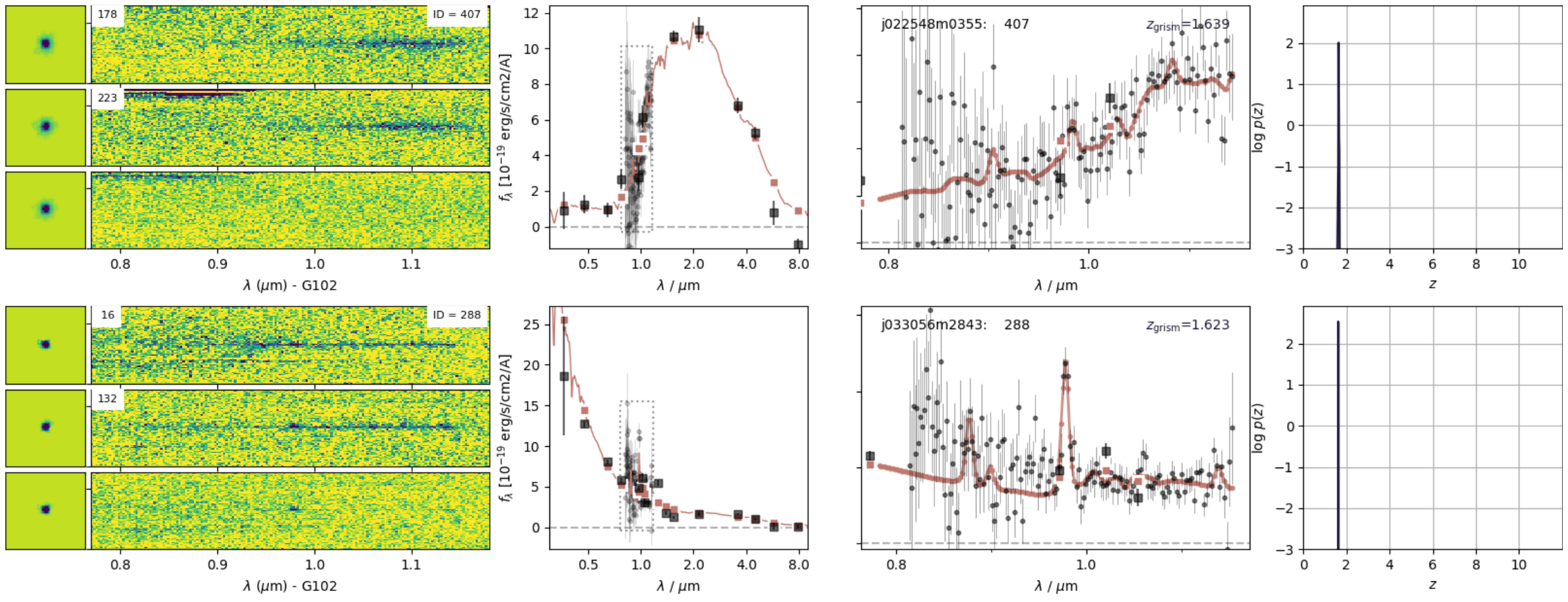}
    \caption{Examples of grism spectra and \texttt{grizli} outputs for two different cluster galaxies. The leftmost panels show the extracted 2D grism spectra from two orients (top and middle), as well as the residual of both with the galaxy continuum and contamination removed (bottom). Object IDs are given in the upper right of the top panel. Numbers in the upper left of the top and middle panels denote the orient of each grism exposure.
    The remaining panels from left-to-right show: the galaxy SED (red line) with photometry (black squares); the 1D grism spectrum; and the probability distribution of $z_\textrm{grism}$.
    }
    \label{fig:grizli_example}
\end{figure*}

We show two examples of cluster members with successful grism spectra and fits in Figure \ref{fig:grizli_example}, alongside their corresponding redshift probabilities.
The spectrum of Object 407 shows a clear 4000\AA\ break, while Object 288 has an [O\textsc{ii}] line.
We note that Object 288 has a difference of only 0.002 between its spectroscopic and grism redshift estimates, while Object 407 had no spectroscopic redshift measurements prior to grism reduction.

\subsubsection{Completeness Limits}

\begin{figure*}[tb]
    \includegraphics[width=1\linewidth]{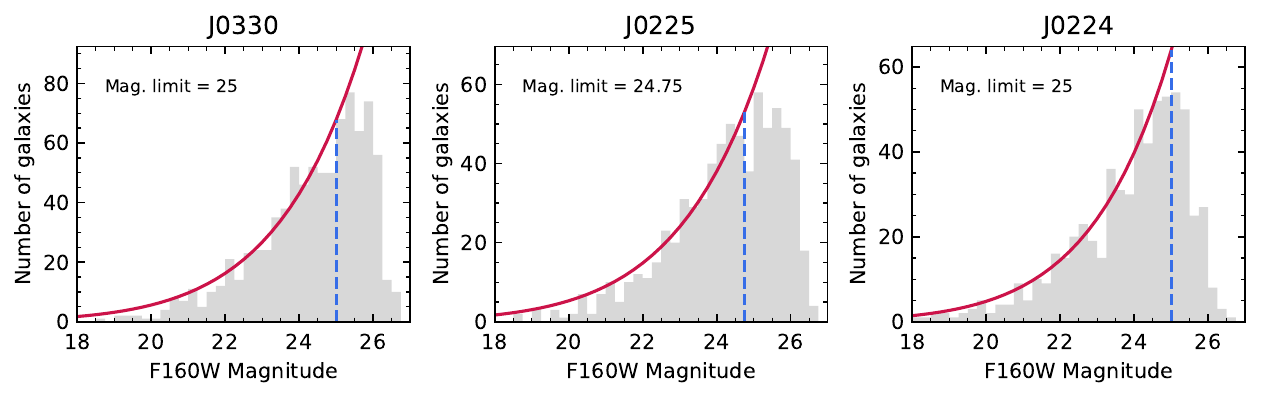}
    \caption{Histograms showing the number density of galaxies from each cluster catalog as a function of F160W magnitude. Solid red lines show a power law fit to the histogram data. We use the difference between the power law fit and the histogram data to determine the 80 percent magnitude limit for each cluster. The magnitude limits are shown as dashed blue lines.}
    \label{fig:obj_mags}
\end{figure*}

For each cluster, we plot a histogram of the F160W magnitudes of all galaxies in the photometric catalog.
We fit a power law to the histogram data, excluding the faintest magnitudes where the number count of objects begins to decline.
Assuming the power law is an estimate of the true number counts of all galaxies, we compare this fit to the histogram data to determine the magnitude at which approximately 80 percent of galaxies are detected.
We repeat this method with varied histogram bin sizes to obtain a more accurate estimate of the 80 percent magnitude limit for each cluster.
An example of this is shown in Figure \ref{fig:obj_mags} along with the magnitude limits determined for each cluster.

Next, we convert the magnitude limits into corresponding stellar mass limits.
In each cluster, we collect all galaxies which have photometric redshifts within $1.4 < z_\textrm{phot} < 1.8$ and are approximately 1 magnitude brighter than the magnitude limit, where we are confident that our sample is complete.
In order to collect a reasonable number of galaxies, we select within a range of $\pm 0.2$ magnitudes.
We set $M_{*,90}$ to be the 90th percentile in stellar mass of these objects (i.e. where 90 percent of these galaxies have masses below $M_{*,90}$). By assuming a constant mass-to-light ratio (M/L), we then use equation \ref{eqn:mass_completeness} to calculate $M_{*, \textrm{lim}}$, the equivalent stellar mass limit at our 80 percent magnitude limit.
This is a reasonable assumption for our purposes, as the largest stellar M/L should not change significantly over 1 magnitude \citep{Marchesini2009}.

\begin{equation}
    M_{*, \textrm{lim}} = M_{*, 90} \times 10^{-0.4}
    \label{eqn:mass_completeness}
\end{equation}

We find stellar mass limits of $10^{9.47} M_{\odot}$ in J0330, $10^{9.50} M_{\odot}$ in J0225, and $10^{9.58} M_{\odot}$ in J0224. Since we will be combining the data from all clusters together, we take the highest mass limit of the three to be our overall mass limit.
For the field data, \citet{Tal2014} determine that the CANDELS fields are 90 percent complete down to stellar masses of $10^{9.04} M_\odot$ for $1.2<z<1.8$.
We therefore assume our cluster and field samples are at least 80 percent complete for stellar masses above $10^{9.58} M_{\odot}$.


\subsubsection{Galaxy Redshift Estimates}
\label{subsec:best_redshift}

For all objects in the photometric cluster catalogs, we determine the source of the best redshift estimate, $z_\textrm{best}$, in order of measurement accuracy.
We prioritize spectroscopic redshift measurements ($z_\textrm{spec}$), followed by reliable grism redshift estimates ($z_\textrm{grism}$, as discussed in Section \ref{subsec:grism}), and finally photometric redshift estimates ($z_\textrm{phot}$) when the other two are not available.
Going forward, we use $z_\textrm{best}$ for sample selection (Section \ref{subsec:cluster_members} and \ref{subsec:field_members}) and calculating projected local densities (Section \ref{subsec:local_density}).


\subsubsection{Cluster Galaxy Selection}
\label{subsec:cluster_members}

When selecting potential cluster members, we require that a galaxy have:
\begin{enumerate}
    \item a non-zero photometric redshift estimate, to help select galaxies with reliable photometry;
    \item a measured flux in F160W, to ensure that the galaxy is reliably detected within the primary imaging;
    \item a stellar mass greater than our mass completeness limit of $10^{9.58} M_{\odot}$.
\end{enumerate}

Galaxies selected as cluster members must also lie within the redshift range of the cluster, based on the source of their best redshift estimate, $z_\textrm{best}$, as outlined below.
We use different selection ranges depending on the source of $z_\textrm{best}$, as the size of typical uncertainties can vary greatly depending on the method used to estimate galaxy redshift. We refer the reader to \citet{vanderBurg2013} and \citet{Matharu2019} who assess the typical rates of false-positive and false-negative cluster members based on different methods of redshift estimation at $z\sim1$.

We use the following formulae to identify cluster members based on the source of their best redshift estimate.
We define $z_\textrm{cl}$ to be the redshift of the cluster, given by the spectroscopic redshift of the brightest cluster galaxy (BCG).
When selecting cluster members based on their spectroscopic redshift (i.e. $z_\textrm{best} = z_\textrm{spec}$), we use the selection threshold from \citet{Muzzin2012}:

\begin{equation}
    \Delta z_\textrm{spec} = \left| \frac{z_\textrm{spec} - z_\textrm{cl}}{1 + z_\textrm{cl}} \right| \leq 0.005
    \label{eqn:z_spec}
\end{equation}

We select cluster members based on their grism redshift, $z_\textrm{grism}$, using the selection threshold found by \citet{Matharu2019}:

\begin{equation}
    \Delta z_\textrm{grism} = \left| \frac{z_\textrm{grism} - z_\textrm{cl}}{1 + z_\textrm{cl}} \right| \leq 0.02
    \label{eqn:z_grism}
\end{equation}

Finally, we select cluster members based on the photometric redshift, $z_\textrm{phot}$, using the selection thresholds in \citet{vanderBurg2013} and \citet{Nantais2016}:

\begin{equation}
    \Delta z_\textrm{phot} = \left| \frac{z_\textrm{phot} - z_\textrm{cl}}{1 + z_\textrm{cl}} \right| \leq 0.05
    \label{eqn:z_phot}
\end{equation}

The number of galaxies selected from each cluster are given in Table \ref{tab:members_by_redshift}. We find a total of 120 cluster members that meet our selection criteria across the three clusters, including 44 based on $z_\textrm{spec}$, 15 based on $z_\textrm{grism}$, and 61 based on $z_\textrm{phot}$. Galaxies which are not chosen as cluster members may still be kept within the sample as potential neighbors for calculating local density (Section \ref{subsec:local_density}).

\begin{deluxetable}{lrrrr}
    \tabletypesize{\footnotesize}
    \tablecolumns{5}
    \tablecaption{Galaxy selection by redshift source.\label{tab:members_by_redshift}}
    \tablehead{
        \colhead{Sample} & \colhead{$z_\textrm{spec}$} & \colhead{$z_\textrm{grism}$} & \colhead{$z_\textrm{phot}$} & \colhead{Total}
    }
    \startdata
        SpARCS J0330&  5 (5)&  6 (5)&  31 (31)& 42 (41)\\
        SpARCS J0225&  14 (13)&  4 (4)&  16 (12)& 34 (29)\\
        SpARCS J0224&  25 (22)&  5 (5)&  14 (14)& 44 (41)\\
        \hline
        COSMOS&  0\tablenotemark{a} &  128 (101)&  220 (159)& 348 (260)\\
        GOODS-S&  15 (7)&  140 (88)&  151 (103)& 306 (198)\\
    \enddata
    \tablecomments{Number of galaxies selected from each cluster and field, sorted by source of best redshift estimate. Selection criteria are described in Sections \ref{subsec:cluster_members} and \ref{subsec:field_members}. Numbers in brackets are the final galaxy selection used in our analysis, including Sérsic index measurements.}
    \tablenotetext{a}{Within the range $1.58103 \leq z_\textrm{spec} \leq 1.646165$.}
\end{deluxetable}

\subsection{3D-HST Fields}
\label{subsec:fields}

We construct a comparative field sample using galaxies from the COSMOS and GOODS-S fields of 3D-HST/CANDELS \citep{Grogin2011,Koekemoer2011}.
These fields were chosen for their wealth of additional data and photometric filter coverage. We use the 3D-HST catalogs for most properties of our field galaxies, including sky positions, F160W fluxes, stellar masses, and rest-frame colors \citep{Brammer2012, Skelton2014}.

We note that the CANDELS/COSMOS field has no significant overdensities at the range of our study \citep{Chiang2014}, however, the GOODS-S field has a known overdensity at $z\sim1.6$ \citep{Castellano2007}.
While we do not exclude galaxies from the GOODS-S overdensity from our field sample, we verify that the trends from the GOODS-S sample are consistent with those of the COSMOS sample, confirming that this does not impact our results.


\subsubsection{Field Galaxy Selection}
\label{subsec:field_members}

We construct a sample of galaxies from the COSMOS and GOODS-S catalogs by following criteria largely equivalent to our cluster member selection, as in Section \ref{subsec:cluster_members}.
As in our cluster member selection, we begin by requiring that all potential field galaxies have non-zero photometric redshift estimates, measured fluxes in F160W, and stellar masses greater than $10^{9.58} M_{\odot}$.
From there, we select field galaxies which lie within a similar redshift range to our clusters, as described below.

Since our clusters are close in redshift, we simply require that the best redshift estimate of each field galaxy, $z_\textrm{best}$, fall within the upper and lower redshift bounds of our three clusters.
We use $z_\textrm{best}$ from the 3D-HST linematched grism catalogs \citep{Brammer2012,Momcheva2016}, and record whether the source of the measurement is ground-based spectroscopy ($z_\textrm{spec}$), WFC3 G141 grism ($z_\textrm{grism}$), or photometry ($z_\textrm{phot}$).
We then use the following ranges to select field galaxies depending on the source of $z_\textrm{best}$:

\begin{align}
    1.58103 & \leq z_\textrm{spec} \leq 1.646165 \\
    1.54212 & \leq z_\textrm{grism} \leq 1.68566 \\
    1.4643 & \leq z_\textrm{phot} \leq 1.76465
\end{align}

The lower bounds of these ranges are found by taking the lower-most bounds of cluster membership set by $z_\textrm{cl} = 1.594$ (J0225), while the upper bounds are found by taking the upper-most bounds of cluster membership set by $z_\textrm{cl} = 1.633$ (J0224), as per equations \ref{eqn:z_spec} through \ref{eqn:z_phot}.

The number of field galaxies selected from COSMOS and GOODS-S are given in Table \ref{tab:members_by_redshift}. Through this process, we select a total of 642 field galaxies, including 15 based on $z_\textrm{spec}$, 260 based on $z_\textrm{grism}$, and 367 based on $z_\textrm{phot}$. As with our cluster samples, we keep non-selected field galaxies as potential neighbors.


\section{Analysis}
\label{sec:analysis}


\subsection{Quantifying Galaxy Structure}
\label{subsec:structure}

\subsubsection{Sérsic Profiles}
\label{subsec:sersic}

A Sérsic profile can be used to represent the intensity of light, $I$, as a function of radial distance from the center of a galaxy, $R$. It is expressed by the following equation:

\begin{equation}
    I(R) = I_e \exp \left\{ -b_n \left[ \left( \frac{R}{R_e} \right)^{1/n} -1 \right] \right\}
    \label{eqn:sersic_profile}
\end{equation}

where $n$ is the Sérsic index, $b_n$ is a function of $n$ such that $R_e$ is the radius at which half of the total light is contained, and $I_e$ is the intensity of light at $R_e$.

We quantify the structure of each galaxy in our sample by its Sérsic index, $n$, and use this as a proxy for morphology.
While typical values of $n$ are found to change with cosmic time, particularly for high-mass quiescent galaxies, studies such as \citet{Buitrago2013} show that a delineation at $n = 2.5$ decently separates disk-dominated morphologies from bulge-dominated ones up to $z \sim 2.5$.
Building on examples set by previous work for $z \gtrsim 1$ (e.g. \citealp{Matharu2019}, \citealp{vanderWel2014}, and particularly \citealp{Strazzullo2023}), we define galaxies with $n<1.5$ as disk-like, galaxies with $n>2.5$ as bulge-like, and galaxies with $1.5<n<2.5$ as intermediate.
We note that the intermediate classification does not necessarily represent a unique morphological class, but does aid in reducing contamination between disk-like and bulge-like classifications.


\subsubsection{Structural Parameters of Field Galaxies}

We use Sérsic profile fits from \citet{vanderWel2012} for our COSMOS and GOODS-S field galaxies.
These fits were performed on the F160W CANDELS mosaics, making them for ideal comparison to Sérsic fits of our cluster data.
We only use Sérsic parameters for galaxies which have a quality flag of 0, indicating good and reliable fits. Galaxies without good fits are still kept in the field sample as possible neighbors, however they are not included in morphological analysis.


\subsubsection{Cluster PSF Construction}
\label{subsec:psfs}

We use cutouts of clean, non-saturated stars as empirical PSFs when modeling the light profiles of our cluster galaxies. Prior to searching for stars that enable the construction of good quality PSFs, we perform source detection on the F160W images of our clusters using Source Extractor \citep{Bertin1996}. We use parameters from the output catalog to locate suitable stars, assist in making object cutouts, and set initial parameters necessary to model Sérsic fits. We also use Source Extractor to create a segmentation map which is used for object masking.

Following \citet{Matharu2019}, we select PSF candidates from objects with $15 < \texttt{MAG\_AUTO} < 19$ and $\texttt{CLASS\_STAR} > 0.5$. The magnitude limits are chosen to exclude oversaturated stars and ensure that the PSF is not too faint. These limits return objects with typical values of $\texttt{FLUX\_RADIUS} \approx 2.5$ and $\texttt{CLASS\_STAR} > 0.8$, showing that this range is, on-average, well-suited to selecting stars.

We create $100 \times 100$ pixel cutouts of each PSF candidate and visually assess them to ensure they are centered on star-like objects.
When assessing the cutouts, we remove candidates which appear dim or contaminated by other objects and extended light sources.
We find four PSF candidates in J0330 and three in J0225. We remove one candidate from each cluster due to contamination by extended objects and galaxies within the cutout image.

In J0224, we detect only a single PSF candidate within our initial selection limits. This candidate is close to our bright magnitude limit with $\texttt{MAG\_AUTO} = 15.20$, and visual inspection shows it is likely an oversaturated star with diffraction spikes that extend beyond the boundaries of the cutout.

We repeat our search in J0224 by extending to dimmer magnitudes and lower \texttt{CLASS\_STAR} values. We find an additional three PSF candidates: one star with $\texttt{MAG\_AUTO} = 19.76$, and two faint stars with $\texttt{MAG\_AUTO} = 21.06$ and $20.55$ respectively. While brighter, the $\texttt{MAG\_AUTO} = 19.76$ star lies too close to a background galaxy to easily mask out, so we choose to eliminate it. Instead, we construct a new PSF by stacking the two faintest stars, utilizing the segmentation map to mask out contamination prior to stacking.

Finally, we test fit a sample galaxy from each cluster to ensure consistency between our PSFs. We compare the output \textsc{Galfit} parameters as well as visually assess the models and residuals to look for significant differences among the PSFs. In the case of J0224, we compare the fits between multiple cluster galaxies using the oversaturated PSF candidate and our stacked PSF. The oversaturated PSF candidate does a poor job of modeling many galaxies, leaving prominent outer rings in the residuals of multiple fits, while our stacked PSF appears to adequately model galaxies with no significant features left in the residuals. We therefore continue our analysis with the three chosen PSFs in J0330, two in J0225, and the stacked PSF in J0224.


\subsubsection{Structural Parameters of Cluster Galaxies with GALFIT}
\label{subsec:galfit}

We fit our cluster members with single-component Sérsic profiles using \textsc{Galfit} \citep{Peng2002, Peng2010a}.
We use a Python-based \textsc{Galfit} wrapper designed by \citet{Matharu2019} to automatically fit batches of images. The wrapper uses a two-stage iterative process to prevent \textsc{Galfit} from becoming stuck on a ``local minimum'' fit, and has been shown to produce results in good agreement with the Sérsic fits of \citet{vanderWel2012}.
We create sigma images for each cluster, following $\sigma = 1/\sqrt{weight}$, rather than leaving \textsc{Galfit} to estimate the noise in each image cutout.

The first stage of our \textsc{Galfit} process involves attempting an initial light profile fit to our galaxies.
We use parameters from the Source Extractor catalog as initial guesses for the integrated magnitude, effective radius ($R_e$), axis ratio, and position angle of each object.
We then set an initial guess of $n=2.5$ for the Sérsic index.
We create image cutouts centered on each primary galaxy to be fit, with a cutout size of $10 \times \texttt{FLUX\_RAD}$, setting a minimum size of 40 pixels and a maximum of 400 pixels.

We simultaneously fit all objects with $\texttt{MAG\_AUTO}<26$ whose centers lie inside the cutout image, using the segmentation map to mask out all other objects.
In the first stage, we also use constraint files to ensure that \textsc{Galfit} sticks to realistic values for galaxy parameters and fits the intended objects.
Among the constraints, we limit the Sérsic index to $0.3 \leq n \leq 8$, following \citet{vanderWel2012}.
Inclusion of both the constraint file and object mask improves the probability of \textsc{Galfit} converging to a solution.

In the second stage, we use the results of the first fit as initial guesses for the final fit.
We extend our image cutouts to $15 \times \texttt{FLUX\_RAD}$, and fix the initial values for the primary object's centroid, axis ratio, and position angle to the output from the first fit, keeping the initial values for other parameters unchanged.
We simultaneously fit secondary objects within the cutout in the same way as the first stage.
If a secondary object was fit independently as a primary galaxy, we use the values found from its primary fit as new initial parameters.
As in the first fit, we mask out all dim objects and objects whose centers lie beyond the edge of the new cutouts. We do not use constraint files in the final fit, and instead allow objects to be fit freely.

\begin{figure}[tb]
    \includegraphics[width=1\linewidth]{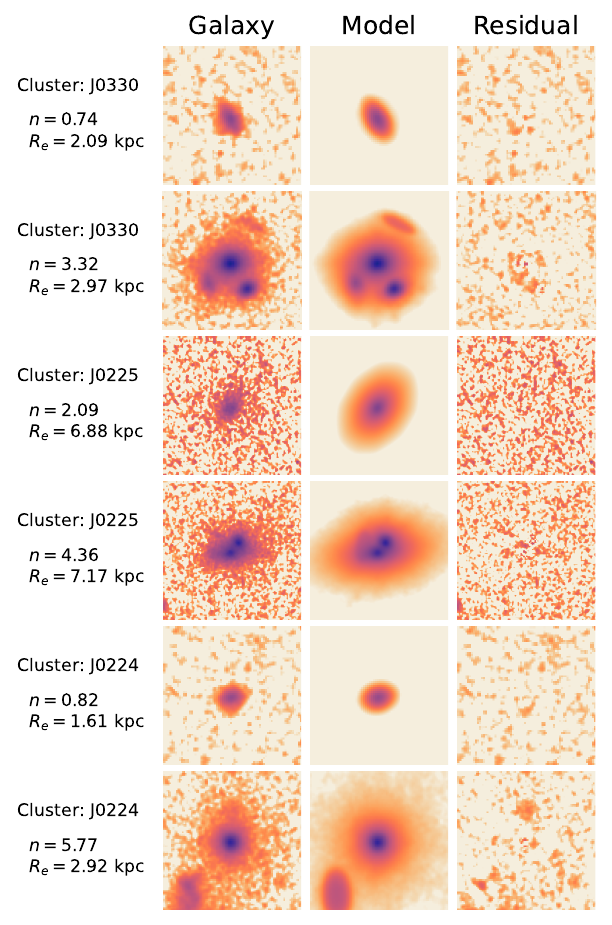}
    \caption{For a handful of cluster members, we show the F160W cutout image, the final \textsc{Galfit} model, and the residual left by subtracting the model from the cutout. Each cutout has a size of $15 \times \texttt{FLUX\_RAD}$ and is centered on the primary galaxy.
    To the left of each row, we list the cluster each galaxy belongs to, as well as the Sérsic index ($n$) and effective radius ($R_e$) of the final model.
    }
    \label{fig:galfit}
\end{figure}

We show examples of the final fits from each cluster in Figure \ref{fig:galfit}.
Most galaxies in our clusters are well-fit with a single Sérsic profile---however, we note that there are a few galaxies where the residuals display a sharp peak in the center and/or a thin ring around the core. We do not comment on the origin of these features, nor do we attempt to modify our fits to specifically accommodate this small subset of objects.

We remove any galaxies from our cluster member sample for which \textsc{Galfit} failed to converge in the second stage or converged with a large reduced $\chi^2$ value, but keep them as potential neighbors for calculating local density. This removes $7.5$ percent of our redshift-selected F160W cluster members overall. J0225 is the most strongly impacted with $5/34$ galaxies removed. The final number of galaxies remaining in each cluster sample after this process is given in brackets in Table \ref{tab:members_by_redshift}.

\subsection{Local Galaxy Density}
\label{subsec:local_density}

\subsubsection{Nth-Nearest Neighbors}

We calculate the local galaxy density of cluster members and field galaxies using the $N$th-Nearest Neighbors method, a convention set by previous major studies (e.g. \citealp{Dressler1980,Dressler1997,Postman2005}). $N$th-Nearest Neighbors calculates the unique local density of any galaxy from the area (or volume) containing the primary galaxy and its $N$ nearest neighboring galaxies. The method requires no prior assumptions about galaxy distribution, and is particularly advantageous for our study as it can be applied to galaxies in any environment, including low-density fields and high-density clusters.

We use $N=5$ to account for the fewer number of cluster members detected at $z \sim 1.6$ compared to low-$z$ ($z\lesssim1$) clusters.
Many observational papers have noted that the choice of $N$ within $\pm$ 2-3 does not significantly influence their results (e.g. \citealp{Postman2005, Bluck2019}). This is echoed by \citet{Cooper2005}, who comment that while increasing $N$ ultimately smooths the density distribution, low-density environments are relatively insensitive to small choices of $N$, while high-density environments are not likely to be sensitive to $N$ so long as it is smaller than the overall richness of the cluster under consideration.

We calculate projected 2D galaxy densities following the methods outlined in \citet{Cooper2005}. While it is possible to calculate 3D galaxy densities, we note that projected densities are less sensitive to uncertainties in line-of-sight velocity/redshift estimates. Given that our samples are not spectroscopically complete, using 2D projected densities ensures more robust measurements for our samples overall.

We use the following equation to calculate the projected local galaxy density, $\Sigma_N$, of a galaxy in a circular area:

\begin{equation}
    \Sigma_N = \frac{N}{\pi {D_N}^2}
    \label{eqn:density}
\end{equation}

where we set $N=5$, and therefore $D_{N} = D_{5}$, the projected distance to the galaxy's 5th nearest neighbor in proper megaparsecs.


\subsubsection{Neighbor Selection}

We eliminate foreground and background interlopers from influencing density calculations by locating neighbors within limited slices in line-of-sight velocity/redshift space. We consider this to be more accurate than traditional background subtraction methods, although we acknowledge that it is only possible due to our sample's extensive multi-band photometry and spectroscopic coverage.

For each primary galaxy (cluster member or field galaxy), we locate all neighboring galaxies that:
\begin{enumerate}
    \item have a non-zero photometric redshift;
    \item have a measured flux in F160W;
    \item have $M_* \geq 10^{9.58} M_{\odot}$;
    \item are within $\pm 3000$ km s$^{-1}$ line-of-sight velocity, calculated using $z_\textrm{best}$. At $z=1.6$, this is equivalent to a redshift range of $\pm 0.026$.
\end{enumerate}

We calculate the projected 2D distance to each neighbor in angular sky coordinates, and then convert to proper megaparsecs using the redshift of the primary galaxy. Neighbors are sorted by increasing distance, and finally $\Sigma_5$ is calculated using equation \ref{eqn:density}.

In rare cases (6 cluster galaxies), we are unable to locate 5 neighbors within $\Delta v=\pm 3000$ km s$^{-1}$ that are detected in F160W. In these cases, we decrease $N$ and use the most distant neighbor found to calculate $\Sigma_N$.


\subsubsection{Edge Effects}

Edge effects can occur when a galaxy is closer to the edge of the detection image than it is to its $N$th neighbor. This introduces a new source of uncertainty in local density measurements, as it cannot be known whether a closer undetected neighbor exists beyond the bounds of the image.

We use the following method to mitigate edge effects in our sample.
First, we determine the edges of each cluster and field image from the F160W weight maps. We then draw out a circle with radius $D_N$ (where $N\leq5$ depending on the most distant neighbor found) centered on each primary cluster member or field galaxy, and calculate the fraction of this circle's area which lies beyond the edge of the detection image ($f_\textrm{edge}$).
For galaxies with $f_\textrm{edge}>0.01$, we introduce a correction factor of $(1-f_\textrm{edge})^{-1}$ to our $\Sigma_N$ calculations, effectively reducing the area used to calculate local density.


\section{Results}
\label{sec:results}


\subsection{The Morphology-Density Relation}
\label{subsec:morph-dens}

We use Sérsic index as a proxy for galaxy morphology following the classifications defined at the end of Section \ref{subsec:sersic}. For ease of discussion, we will refer to our disk-like and bulge-like classifications as simply ``disk'' and ``bulge'' throughout this section.

\begin{figure*}[htb]
    \includegraphics[width=1\linewidth]{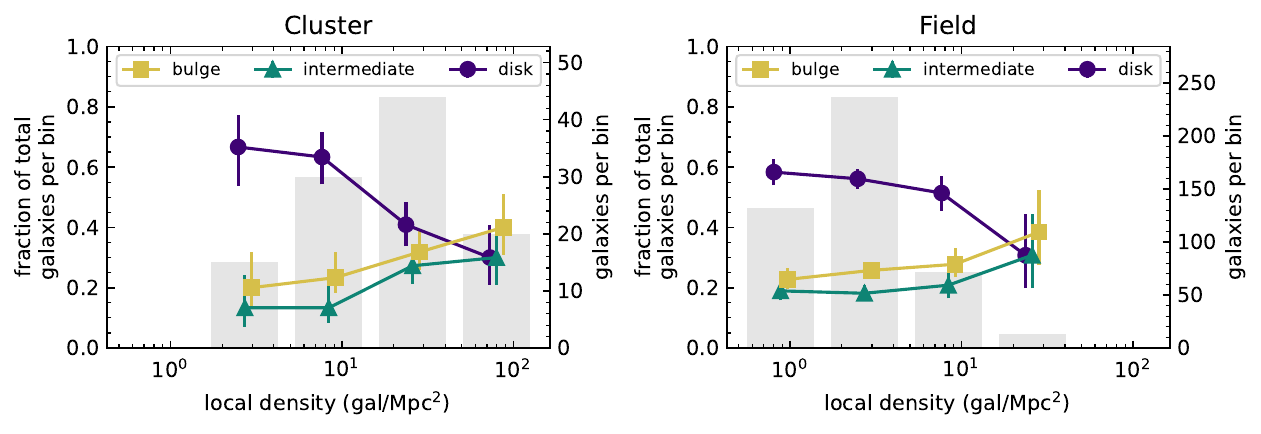}
    \caption{The binned morphology-density relation in cluster galaxies (left) and field galaxies (right) at $z\sim1.6$. The fraction of bulge-like galaxies are shown as yellow squares, intermediate galaxies as green triangles, and disk-like galaxies as purple circles. Error bars represent the 68 percent credible interval. The number of galaxies in each bin is underplotted as a grey histogram, following the right-hand $y$-axis.}
    \label{fig:MDR_cl+fd_binned}
\end{figure*}

In Figure \ref{fig:MDR_cl+fd_binned}, we plot the morphological fractions of cluster members (left) and field galaxies (right) in bins of increasing local density.
Points are offset from bin centers for clarity, with disk galaxies ($n<1.5$) plotted as purple circles, intermediate galaxies ($1.5\leq n\leq 2.5$) as green triangles, and bulge galaxies ($n>2.5$) as yellow squares.
For ease of comparison between cluster and field, we plot both samples using identical local density bins. In either case, we cut off the lowest or highest density bins where there are fewer than 10 galaxies per bin or less than one galaxy of any morphological type. The number of galaxies in each bin is underplotted as a grey histogram, with values given by the right-hand $y$-axis.
Error bars represent the 68 percent credible interval on the fraction, estimated using Jeffreys Bayesian credible interval for binomial proportions.

The binned trends as in Figure \ref{fig:MDR_cl+fd_binned} follow the traditional presentation of results in studies of the morphology-density relation. As an alternative to traditional binning, we also present our results using a fixed-width box kernel, which shares similarities with statistical methods of calculating rolling means/medians. This has the benefit of making our results less sensitive to specific binning choices, while also allowing us to extrapolate more detail with the same sample size. This technique will be used throughout the paper, and we will indicate the width and spacing of the box kernel for each plot.

\begin{figure*}[htb]
    \includegraphics[width=1\linewidth]{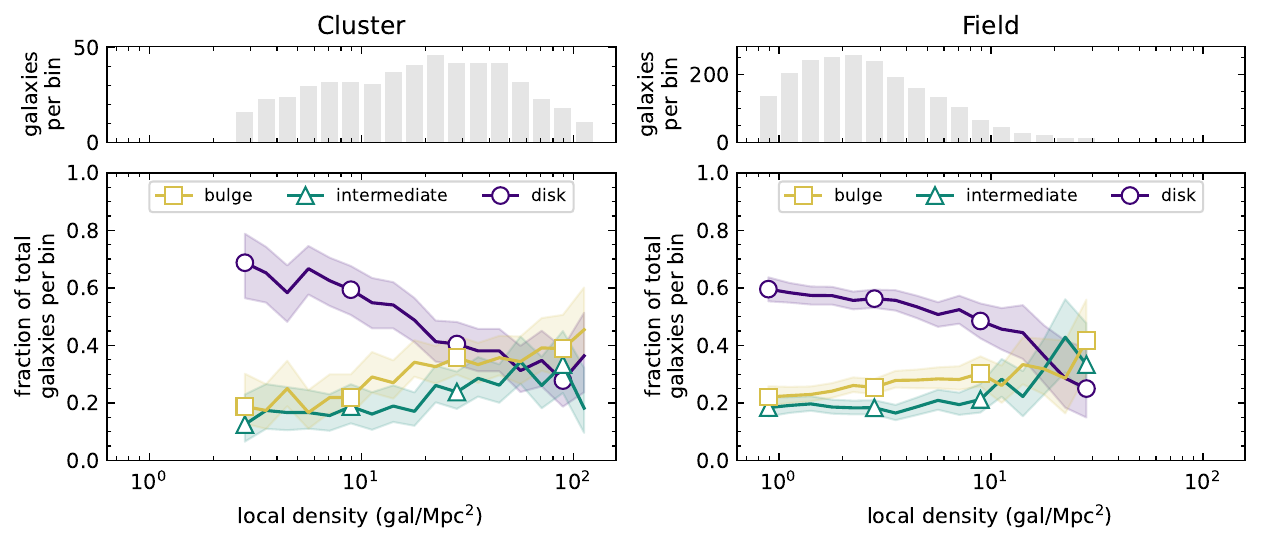}
    \caption{The morphology-density relation in cluster galaxies (left) and field galaxies (right) at $z\sim1.6$, plotted using a fixed-width box kernel as described in Section \ref{subsec:morph-dens}. The fraction of bulge-like galaxies is shown as a yellow line with square symbols, intermediate galaxies as a green line with triangles, and disk-like galaxies as a purple line with circles. Symbols are placed along each line showing the closest distance between fully independent bins. Shaded regions represent the 68 percent credible interval. Upper panels show the number of galaxies in each bin (not representative of bin size).}
    \label{fig:MDR_cl+fd_kernel}
\end{figure*}

In Figure \ref{fig:MDR_cl+fd_kernel}, we plot the morphology-density relation of cluster members (left) and field galaxies (right) using a fixed-width box kernel. The box kernel is constructed with bin widths of $\pm0.25\log(\Sigma_N)$ and a spacing of $0.1\log(\Sigma_N)$ between bin centers. As in Figure \ref{fig:MDR_cl+fd_binned}, we use identical density bins for both cluster and field samples, and cut off when the number of galaxies per bin falls below $10$.
Symbols have been added along each line to guide the eye, and are spaced to show the closest distance between two fully independent bins.
Shaded regions show the statistical errors, calculated in the same way as Figure \ref{fig:MDR_cl+fd_binned}. The total number of galaxies per bin is plotted as a histogram in the upper panel of each sample. While the histograms are aligned with the bin centers, they are not representative of the true bin widths.

In Figures \ref{fig:MDR_cl+fd_binned} and \ref{fig:MDR_cl+fd_kernel}, we see a clear trend in the morphology-density relation of both cluster members and field galaxies, such that the fraction of bulge galaxies increases with local density, while the fraction of disk galaxies decreases.
Disk galaxies make up the majority of both samples at low densities (up to $69^{+10}_{-12}$ percent in the cluster, $60^{+4}_{-4}$ percent in the field), while bulge and intermediate fractions are similarly low (roughly 15 to 25 percent).
The intermediate fraction appears to increase along with the bulge fraction, although it remains slightly lower at all densities.
At high densities, the fractions of each morphological type converge, after which the bulge fraction exceeds the disk fraction, reaching a maximum of $46^{+15}_{-9}$ percent in the cluster and $42^{+14}_{-12}$ percent in the field.

It is evident that the morphology-density relation is already in place in clusters by $z\sim1.6$, and follows a trend qualitatively similar to what has been observed in lower redshift studies. We have also found evidence of a morphology-density relation in the field sample at $z\sim1.6$.
Interestingly, both trends also appear fairly similar, despite the samples being drawn from different global environments.

\begin{figure*}[htb]
    \includegraphics[width=1\linewidth]{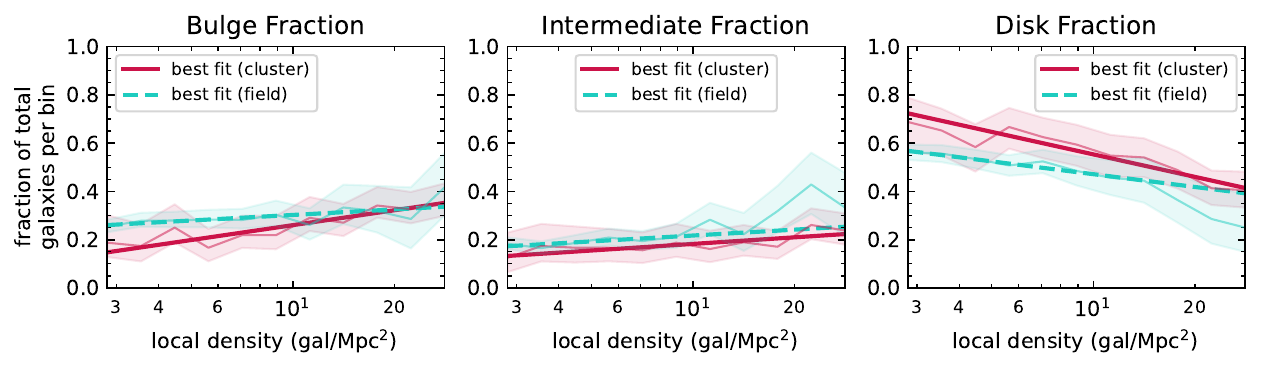}
    \caption{The fraction of bulge-like, intermediate, and disk-like galaxies as a function of projected local galaxy density, $\Sigma_N$, in both the cluster (red) and field sample (cyan). Thin lines and shaded regions show the data plotted using a fixed-width box kernel. Thick lines show the best-fit linear trendlines for cluster galaxies (solid red) and field galaxies (dashed cyan). Parameters for each trendline are given in Table \ref{tab:MDR_slopes}. Cluster and field fractions overall agree within uncertainties, with the field sample perhaps having a lower fraction of disk-like galaxies overall.
    }
    \label{fig:MDR_linearfits}
\end{figure*}

To compare of the strength of the morphology-density relation in each sample, we find the best-fit trendlines to the fractions of each morphological type over the local density range spanned by both cluster and field samples.
This is shown in Figure \ref{fig:MDR_linearfits}, with cluster fractions plotted in red and field fractions plotted in cyan.
We show the comparison between the fraction of bulge galaxies (left), intermediate galaxies (middle), and disk galaxies (right).
The thin lines and shaded regions are the same as in Figure \ref{fig:MDR_cl+fd_kernel}.
Best-fit lines are found using weighted linear regression on the fractions of each morphological type as a function of $\log (\Sigma_N)$.
The resulting trendlines are plotted as thick solid lines (cluster) and thick dashed lines (field).

The parameters for each best-fit line are given in Table \ref{tab:MDR_slopes}.
In both samples, the trendlines indicate a steeper decrease in the fraction of disk galaxies and a shallow increase in the fraction of bulge and intermediate galaxies.
The slopes of the bulge and disk fractions appear shallower in the field sample, although the field sample appears to generally have a lower fraction of disk galaxies than the cluster.
In the field sample, the disk and intermediate fractions also appear to diverge from a linear trend for local densities greater than 10 gal Mpc$^2$, although this is notably the density range with the fewest field galaxies.
Overall, however, the cluster and field fractions are likely consistent within error bars.

\begin{deluxetable*}{l RR c RR}
    \tabletypesize{\small}
    \tablecolumns{5}
    \tablecaption{Linear regression parameters for the best-fit line to each morphological fraction in the morphology-density relation, as plotted in Figure \ref{fig:MDR_linearfits}. \label{tab:MDR_slopes}}
    \tablehead{
        \colhead{} & \multicolumn{2}{c}{Cluster} & \colhead{} & \multicolumn{2}{c}{Field} \\
        \cline{2-3} \cline{5-6}
        \colhead{Fraction} & \colhead{Slope} & \colhead{Zero Point} & \colhead{} & \colhead{Slope} & \colhead{Zero Point}
    }
    \startdata
        Bulge-like& 0.20\pm0.03 & 0.06\pm0.03 && 0.08\pm0.02 & 0.23\pm0.01 \\
        Intermediate& 0.09\pm0.03 & 0.09\pm0.03 && 0.08\pm0.03 & 0.14\pm0.02 \\
        Disk-like& -0.31\pm0.04 & 0.86 \pm 0.04 && -0.18\pm0.03 & 0.65\pm0.02 \\
    \enddata
    \tablecomments{Morphological fractions are given as a function of $\log(\Sigma_N)$.}
\end{deluxetable*}

\begin{figure}[tb]
    \includegraphics[width=1\linewidth]{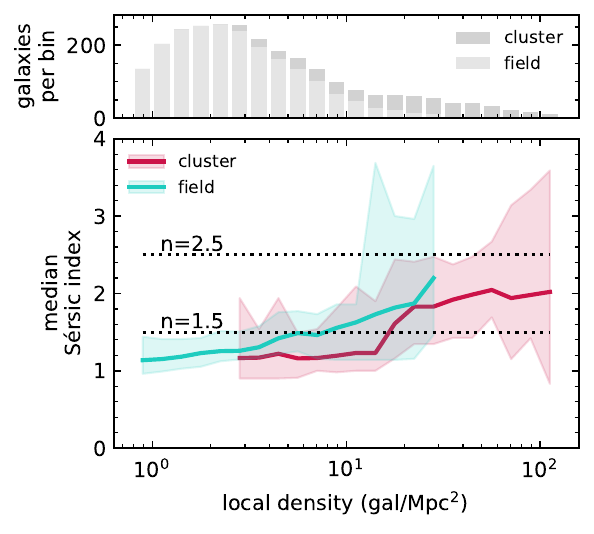}
    \caption{Median Sérsic index as a function of projected local galaxy density, $\Sigma_N$, for cluster galaxies (solid red) and field galaxies (solid cyan), plotted using a fixed-width box kernel.
    Shaded regions show the 95\% confidence interval from bootstrapping. Horizontal dotted lines indicate $n=1.5$ and $n=2.5$. The upper panel shows the number of field galaxies (light grey) and cluster galaxies (darker grey) in each bin as a stacked histogram.
    }
    \label{fig:median_sersics}
\end{figure}

In Figure \ref{fig:median_sersics}, we take advantage of the quantitative nature of Sérsic index measurements and plot the median Sérsic index of galaxies as a function of local density.
We use the median as it is more robust against outliers and skewed distributions than the mean. It is also more physically intuitive to interpret: half of the galaxies in the sample should fall above the median, and half below.
We plot the median using a fixed-width box kernel, as in Figure \ref{fig:MDR_cl+fd_kernel}.
Shaded regions represent the 95 percent confidence interval on the median calculated from bootstrapping.
The number of galaxies in each bin is shown in the upper panel as a stacked histogram.
For comparison, we plot dotted horizontal lines to highlight where the median lies in relation to $n=1.5$ and $n=2.5$.

Figure \ref{fig:median_sersics} can be considered a \textit{structure}-density relation, rather than a true morphology-density relation.
It is not dependent on pre-sorting galaxies into morphological types, and the results are invariant to the choice of $n$ separating disk and bulge classifications.
In both samples, we see that the median Sérsic index increases as a function of local galaxy density.
In the field, the median Sérsic index rises smoothly from a value of $1.14^{+0.30}_{-0.18}$ up to $2.20^{+1.45}_{-0.74}$.
In the cluster, it rises from $1.17^{+0.77}_{-0.27}$ up to a peak value of $2.05^{+0.62}_{-0.36}$, with the bulk of this change occuring at local densities greater than 14 gal Mpc$^{-2}$.
While the median Sérsic index of field galaxies appears higher than that of cluster galaxies (within the range of overlapping density), both trends are broadly consistent within error bars.
These results are also consistent with our observed morphology-density relation (Figures \ref{fig:MDR_cl+fd_binned} and \ref{fig:MDR_cl+fd_kernel}).

As a final piece of analysis, we perform two sets of statistical tests to assess whether the cluster and field trends show significant differences in the distribution of galaxy morphology with local density.
For both tests, we use binned galaxy counts from the three local density bins in Figure \ref{fig:MDR_cl+fd_binned} where both samples have at least 10 total galaxies.
Firstly, we perform one-way chi-square goodness-of-fit tests, assessing the null hypothesis that the cluster trends are consistent with the field trends.
We find an  overall p-value of 0.79, with p-values for individual morphological types ranging from 0.93 for bulge galaxies and 0.91 for intermediate galaxies, down to 0.48 for disk galaxies.
Secondly, in each density bin, we perform a chi-square test of independence to assess the null hypothesis that the cluster and field samples are not independent.
We find p-values ranging from 0.51 to 0.80.
With a minimum p-value of 0.48 across all tests, the results indicate that we cannot reject either null hypothesis.

The observed trends of galaxy morphology with local density appear fairly similar for both the cluster and field samples, despite their different density distributions.
While the morphological fractions in the field appear to converge at slightly lower densities, this still falls within the statistical uncertainty of the cluster.
The statistical tests also indicate that there is a reasonable probability that the cluster and field trends are consistent with each other.
We cannot rule out the possibility that galaxies in both samples follow the \textit{same} morphology-density relation.
If this is indeed the case, it would suggest that the relationship between morphology (or structure) and local galaxy density is independent of global environment at $z\sim1.6$.


\subsubsection{Towards a ``Unified'' Relation}
\label{subsec:unified_sample}

To explore the theory that the morphology-density relation is independent of global environment, we create a simplistic ``unified'' sample of galaxies at $z\sim1.6$ by combining our cluster member and field galaxy samples. All galaxies are weighted equally in the unified sample, and the number of galaxies from each environment are shown as a stacked histogram.
If the theory is sound, then the presentation of this unified sample allows us to take advantage of the high number of low-density field galaxies as well as the greater number of high-density cluster members to provide a better look at the morphology-density relation at $z\sim1.6$.

\begin{figure}[tb]
    \includegraphics[width=1\linewidth]{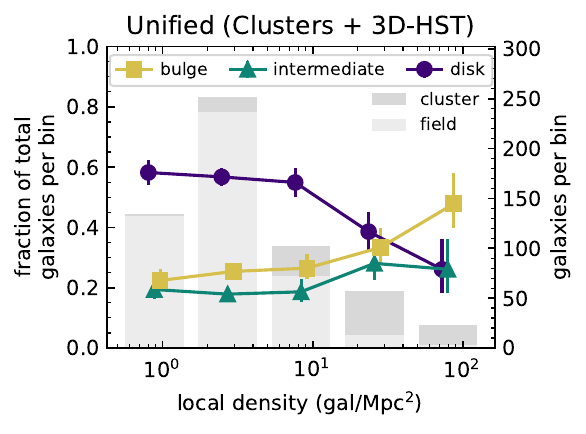}
    \caption{The binned morphology-density relation in the unified sample at $z\sim1.6$. The plot is shown as in Figure \ref{fig:MDR_cl+fd_binned}, with the total number of galaxies in each bin underplotted as a stacked histogram showing the number of field galaxies (light grey) and cluster galaxies (darker grey) combined.}
    \label{fig:MDR_unified_binned}
\end{figure}

\begin{figure}[tb]
    \includegraphics[width=1\linewidth]{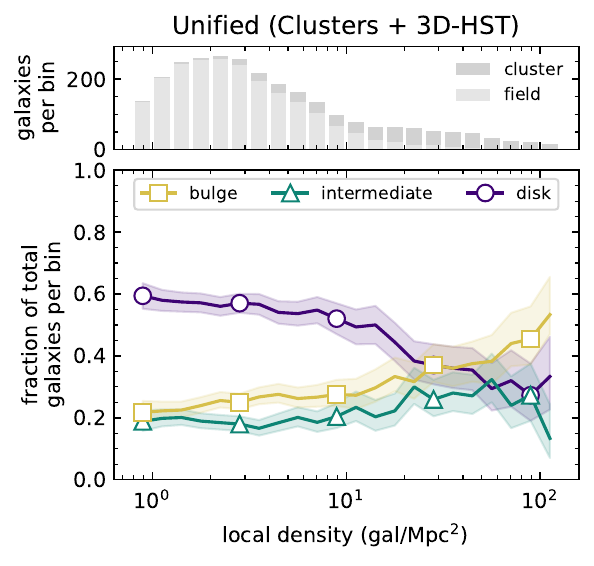}
    \caption{The morphology-density relation in the unified sample at $z\sim1.6$, plotted using a fixed-width box kernel. The plot is shown as in Figure \ref{fig:MDR_cl+fd_kernel}, with the upper panel showing a stacked histogram of the number of field galaxies (light grey) and cluster galaxies (darker grey) combined in each bin. Symbols are once again placed to show the distance between fully independent bins.}
    \label{fig:MDR_unified_kernel}
\end{figure}

We separate our galaxies into morphological classifications as before.
We plot the fraction of each morphological type as a function of local galaxy density for the unified sample in Figure \ref{fig:MDR_unified_binned} with traditional bins, and in Figure \ref{fig:MDR_unified_kernel} with a fixed-width box kernel.
Plot details and bins are the same as in Figure \ref{fig:MDR_cl+fd_binned} and Figure \ref{fig:MDR_cl+fd_kernel}, respectively, with the number of galaxies in each bin plotted as a stacked histogram.

The trend in the morphology-density relation of the unified sample is in good agreement with the separate trends of both the cluster and field samples.
Disk galaxies again make up the majority at low local densities, at $59^{+4}_{-4}$ percent, while bulge and intermediate galaxies make up $22^{+4}_{-3}$ percent and $ 19^{+3}_{-3}$ percent, respectively.
Morphological fractions begin to converge at local densities of $\sim 28$ gal Mpc$^{-2}$.
While the intermediate fraction follows just under the bulge fraction at most densities, the unified relation suggests that it peaks as bulge galaxies become the dominant morphological type.
At high densities, the bulge fraction reaches a maximum of $53^{+12}_{-6}$ percent.

Our results support the idea that the morphology-density relation may be universal at this epoch.
Regardless of global environment, it is clear that a morphology-density relation is already in place by $z\sim1.6$, resulting in a decrease in the fraction of disk galaxies and an increase in the fraction of bulge galaxies as local galaxy density increases. Similarly, the median Sérsic index of galaxies also increases with local galaxy density.
We note that even if the relation is not truly universal, the presence of a clear trend in the field sample opposes the idea that cluster-specific processes are the primary drivers behind the morphology-density relation at this epoch.

\subsection{The Morphology-Mass Relation}
\label{subsec:morph-mass}

As previously mentioned, it is important to explore the relationship between galaxy structure and stellar mass to determine whether galaxy stellar mass plays a significant role in the observed morphology-density relation.

\begin{figure*}[htb]
    \includegraphics[width=1\linewidth]{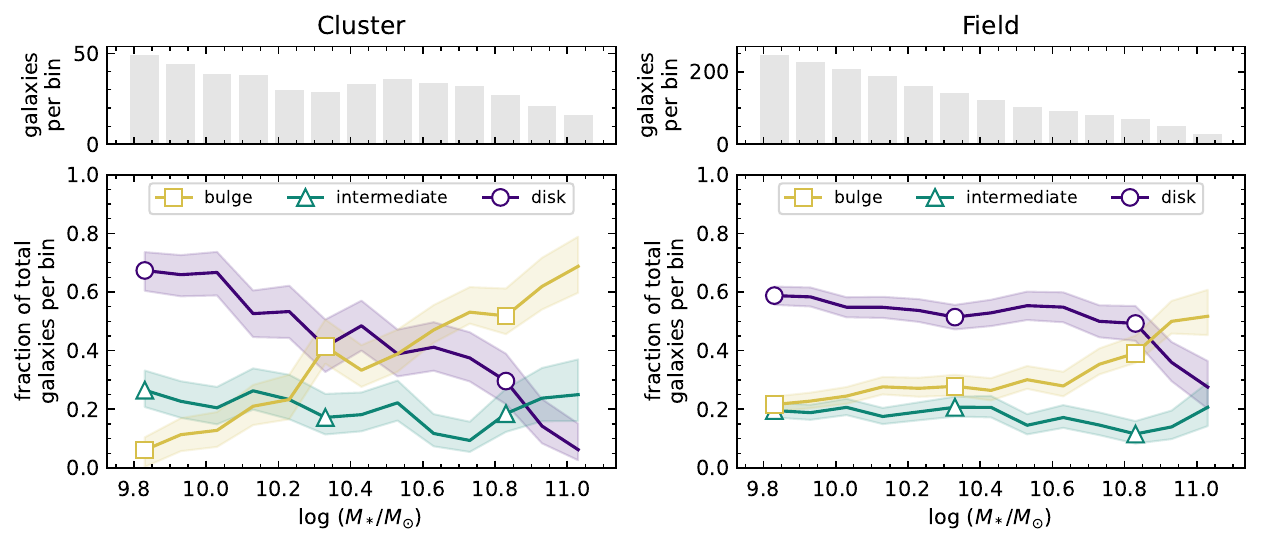}
    \caption{The morphology-mass relation in cluster galaxies (left) and field galaxies (right) at $z\sim1.6$, plotted using a fixed-width box kernel as described in Section \ref{subsec:morph-mass}. The fraction of bulge-like galaxies is shown as a yellow line with square symbols, intermediate galaxies as a green line with triangles, and disk-like galaxies as a purple line with circles. Symbols are placed along each line showing the closest distance between fully independent bins. Shaded regions represent the 68 percent credible interval. Upper panels show the number of galaxies in each bin (not representative of bin widths).}
    \label{fig:MMR_kernel}
\end{figure*}

In Figure \ref{fig:MMR_kernel}, we plot the fraction of each morphological type as a function of galaxy stellar mass for both cluster and field galaxies.
Results are plotted using a fixed-width box kernel with bin sizes of $\pm 0.25\log(M_*/M_{\odot})$ and a spacing of $0.1\log(M_*/M_{\odot})$ between bins. Markers are once again added to guide the eye, and spaced to show the closest distance between two fully independent bins.
The smoothed distribution of galaxies with stellar mass can be seen in the upper panels of Figure \ref{fig:MMR_kernel}, with a histogram representing the number of galaxies in each bin.

In both samples, we see a general trend that the fraction of bulge galaxies increases with stellar mass, while the fraction of disk galaxies decreases proportionally.
However, this effect appears more prominent in the cluster sample than the field.
From the lowest to highest stellar mass ranges probed, the fraction of bulge galaxies in the cluster increases from a low of only $6^{+4}_{-6}$ percent up to $69^{+10}_{-9}$ percent, while the fraction of disk galaxies decreases from $67^{+6}_{-7}$ percent down to $6^{+9}_{-4}$.
This change is less drastic in the field sample, where the fraction of bulge galaxies increases from $22^{+3}_{-2}$ percent to $52^{+9}_{-6}$ percent, and the fraction of disk galaxies decreases from $59^{+3}_{-3}$ percent down to $28^{+9}_{-7}$ percent.
In both samples, the intermediate fraction remains lower across the stellar mass range, with a typical value of $\sim20$ percent in the field and $\sim20$ to 25 percent in the cluster.

\begin{figure*}[htb]
    \includegraphics[width=1\linewidth]{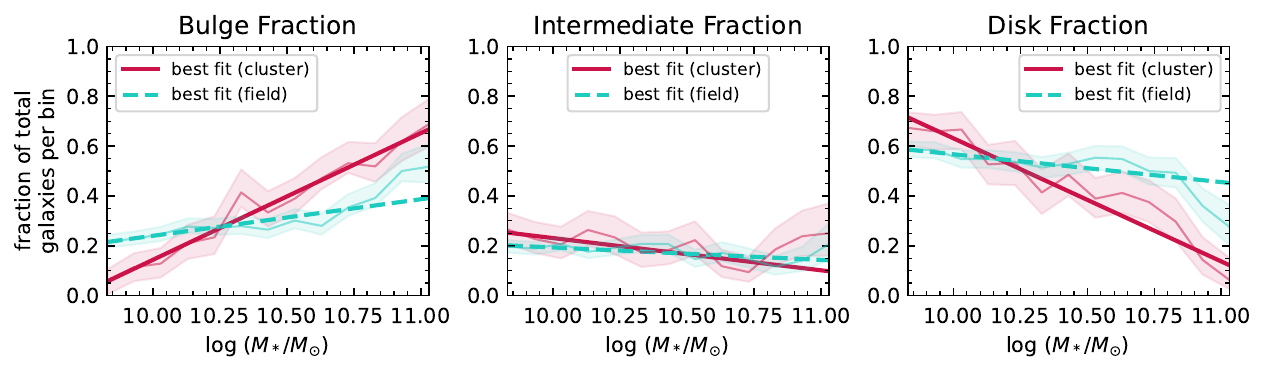}
    \caption{The fraction of bulge-like, intermediate, and disk-like galaxies as a function of galaxy stellar mass, $\log(M_*/M_{\odot})$, in both the cluster (red) and field sample (cyan). Thin lines and shaded regions show the data plotted using a fixed-width box kernel. Thick lines show the best-fit trendlines for cluster galaxies (solid red) and field galaxies (dashed cyan). Parameters for each trendline are given in Table \ref{tab:MMR_slopes}.}
    \label{fig:MMR_linearfits}
\end{figure*}

To better compare the strength of the morphology-mass relation between cluster and field samples, we once again use linear trendlines.
We use weighted linear regression to find the best-fit trendline to each morphological fraction as a function of stellar mass across the whole stellar mass range, shown in Figure \ref{fig:MMR_linearfits}.
Parameters of each best-fit trendline are given in Table \ref{tab:MMR_slopes}, with errors.

\begin{deluxetable*}{l RR c RR}
    \tabletypesize{\footnotesize}
    \tablecolumns{5}
    \tablecaption{Linear regression parameters for the best-fit line to each morphological fraction in the morphology-mass relation, as plotted in Figure \ref{fig:MMR_linearfits}. \label{tab:MMR_slopes}}
    \tablehead{
        \colhead{} & \multicolumn{2}{c}{Cluster} & \colhead{} & \multicolumn{2}{c}{Field} \\
        \cline{2-3} \cline{5-6}
        \colhead{Fraction} & \colhead{Slope} & \colhead{Zero Point} & \colhead{} & \colhead{Slope} & \colhead{Zero Point}
    }
    \startdata
        Bulge-like& 0.51\pm0.02 & -4.9\pm0.2 && 0.15\pm0.02 & -1.2\pm0.2 \\
        Intermediate& -0.13\pm0.04 & 1.5\pm0.4 && -0.05\pm0.02 & 0.7\pm0.2 \\
        Disk-like& -0.49\pm0.04 & 5.6\pm0.4 && -0.11\pm0.03 & 1.7\pm0.3 \\
    \enddata
    \tablecomments{Morphological fractions are given as a function of $\log(M_*/M_{\sun})$.}
\end{deluxetable*}

Figure \ref{fig:MMR_linearfits} highlights the general trends found in Figure \ref{fig:MMR_kernel}.
The slopes of both the bulge and disk trendlines are significantly steeper in the cluster sample compared to the field.
This may suggest that stellar mass has a stronger impact on the fraction of bulge and disk galaxies in clusters.
Additionally, in both samples, the slopes of the trendlines indicate that the increase in the bulge fraction is generally proportional to the decrease in the disk fraction. Meanwhile, the fraction of intermediate galaxies is fairly similar between cluster and field samples, with a flatter, slightly negative slope.
This could indicate a gradual decline in the fraction of intermediate galaxies with stellar mass, although the trend is less consistent.

While the difference in trends may appear significant, we must note two caveats to these results.
Firstly, while the trendlines in Figure \ref{fig:MMR_linearfits} seem to adequately describe most of the data, features in the high-mass end suggest that the morphology-mass relation may deviate from a simple (or singular) linear trend.
For example, the disk fraction in the field sample decreases by only 10 percent from $9.83 \lesssim \log(M_*/M_{\odot}) \lesssim 10.83$ (1 dex in stellar mass), but decreases another 21 percent from $10.83 \lesssim \log(M_*/M_{\odot}) \lesssim 11.03$ (0.2 dex in stellar mass).
Linear trendlines are suitable to make a general relative comparison, however, interpretation of these trendlines and their parameters should be treated with caution.
Secondly, we note that the largest differences between the trends in Figure \ref{fig:MMR_kernel} are the lack of low-mass bulge galaxies and high-mass disk galaxies in the cluster.
These differences can contribute to the steepness of the slopes in the cluster trendlines.
Therefore, the true difference in the trends between cluster and field galaxies may be smaller than what is suggested by the trendlines in Figure \ref{fig:MMR_linearfits}.

\begin{figure}[tb]
    \includegraphics[width=1\linewidth]{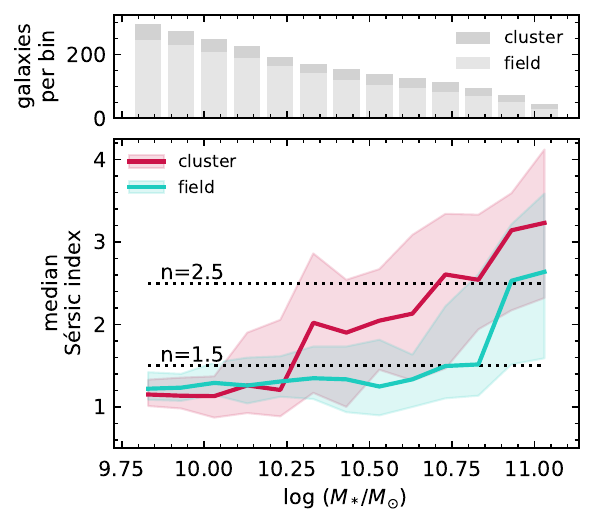}
    \caption{Median Sérsic index as a function of galaxy stellar mass for cluster galaxies (solid red) and field galaxies (solid cyan), plotted using a fixed-width box kernel. Shaded regions show the 95\% confidence interval from bootstrapping. Horizontal dotted lines indicate $n=1.5$ and $n=2.5$. The upper panel shows the number of field galaxies (light grey) and cluster galaxies (darker grey) in each bin as a stacked histogram.
    }
    \label{fig:median_sersic-mass}
\end{figure}

In Figure \ref{fig:median_sersic-mass}, we plot the median Sérsic index of galaxies as a function of stellar mass.
The details of this plot are similar to Figure \ref{fig:median_sersics}, but with identical bins to Figure \ref{fig:MMR_kernel}.
In both samples, the median Sérsic index generally increases with galaxy stellar mass.
At low stellar masses, $\log(M_*/M_\odot)\lesssim10.23$, galaxies in both samples have similar median Sérsic indices.
However, as stellar mass grows, the median Sérsic index of cluster galaxies diverges from a flat trend and increases above that of field galaxies.
Comparatively, field galaxies delay their Sérsic index growth until higher stellar masses.
Across the stellar mass range, the cluster sample ranges from a median Sérsic index of $1.15^{+0.18}_{-0.17}$ up to $3.23^{+0.89}_{-1.28}$, while the field sample ranges from a median Sérsic index of $1.22^{+0.19}_{-0.15}$ up to $2.64^{+0.95}_{-1.05}$.
The similarity of both samples at the low-mass end stands in contrast to the morphological fractions in Figure \ref{fig:MMR_kernel}, suggesting that much of the difference between bulge and disk fractions in the low-mass regime results from the distribution (and not the amount) of galaxies with $n\gtrsim1.2$.

Beyond the general trend, two additional features stand out in Figure \ref{fig:median_sersic-mass}.
The median Sérsic index of cluster galaxies rises above $n=1.5$ at $\log(M_*/M_{\odot}) > 10.23$, indicating that less than half of cluster galaxies would be classified as disks above this mass.
At this same mass, the majority of field galaxies remain disks, and the overall trend in field galaxies remains fairly flat until high masses.
Meanwhile, at $\log(M_*/M_{\odot}) > 10.83$, the median Sérsic index of both cluster and field galaxies rises above $n=2.5$.
This indicates that, regardless of environment, more than half of galaxies at $\log(M_*/M_{\odot}) > 10.83$ would be classified as bulges.
This is in contrast to Figure \ref{fig:median_sersics}, where the median Sérsic index of either individual sample remains below $n=2.5$ for all projected local densities.

As in Section \ref{subsec:morph-dens}, we perform two sets of statistical tests to assess whether there are significant differences in the morphology-mass relation of the cluster and field samples.
For these tests, we use galaxy counts in 4 stellar mass bins of equal width from $9.58 < \log(M_*/M_\odot)<11.25$.
We perform one-way chi-square goodness-of-fit tests to assess whether the cluster trends are consistent with the field, finding an overall p-value of 0.42.
For individual morphological types, we find that bulge galaxies have the lowest p-value with 0.32, while disk and intermediate galaxies have p-values of 0.65 and 0.90, respectively.
We then perform a chi-square test of independence on each stellar mass bin, and find p-values ranging from 0.11 to 0.65.
The lowest p-value occurs in the lowest mass bin ($9.58<\log(M_*/M_\odot)<10.00$), while the highest two mass bins both have p-values of 0.38.
Since all tests give p-values above the nominal critical value of 0.05, we cannot rule out the null hypotheses.
However, the majority of tests indicate less than a 50 percent chance that the cluster and field trends are consistent with each other.

The statistical tests are ultimately inconclusive regarding whether the differences in the morphology-mass relation of cluster and field samples are significant.
Additionally, the uncertainties in Figure \ref{fig:median_sersic-mass} are still quite large.
It is possible that the cluster and field trends may be consistent with each other.
A larger sample of galaxies (particularly in clusters) would be needed to confidently resolve whether this is the case.
In the absence of this, we investigate possible sources of the potential difference in the morphology-mass relations.

\begin{figure*}[tb]
    \includegraphics[width=1\linewidth]{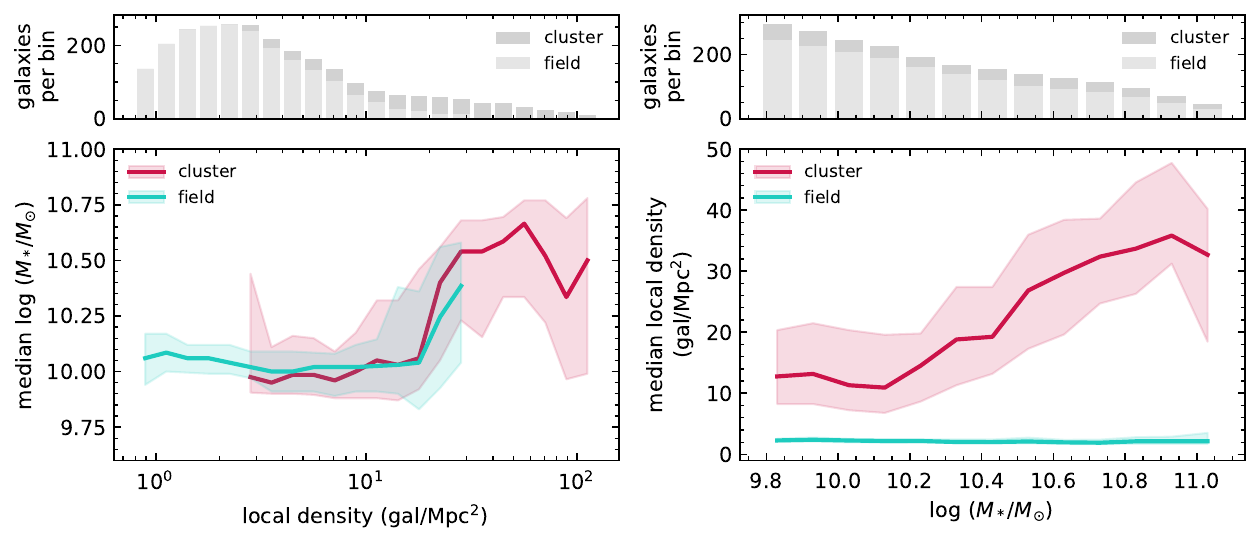}
    \caption{Left: Median stellar mass as a function of projected local galaxy density, $\Sigma_N$. Right: Median local density as a function of galaxy stellar mass, $\log(M_*/M_\odot)$.
    Trends in cluster galaxies are shown in red, while field galaxies are shown in cyan, plotted using fixed-width box kernels.
    Shaded regions show the 95\% confidence interval from bootstrapping. Upper panels show the number of field galaxies (light grey) and cluster galaxies (darker grey) in each bin as a stacked histogram.}
    \label{fig:density-mass}
\end{figure*}

In Figure \ref{fig:density-mass}, we examine the correlation between local environment and galaxy stellar mass in both cluster and field samples.
In the left-hand panels, we plot median stellar mass as a function of projected local galaxy density.
In the right-hand panels, we plot median local density as a function of galaxy stellar mass.
We use identical box kernels and binning as in Figure \ref{fig:median_sersics} and Figure \ref{fig:median_sersic-mass}, for the left and right panels, respectively.
Note that in the right panel, median local density is plotted on a linear scale along the $y$-axis.

In the left panel of Figure \ref{fig:density-mass}, we see that the cluster and field samples appear to follow very similar trends in median stellar mass with local galaxy density.
The median stellar mass of both cluster and field galaxies is fairly constant at $\sim10^{10} M_\odot$ for $\Sigma_N\lesssim18$ gal Mpc$^{-2}$.
Above this density, median stellar mass increases with local density in both samples,
with cluster galaxies peaking at a median stellar mass of $\sim10^{10.67} M_\odot$ at $\Sigma_N\sim56$ gal Mpc$^{-2}$.
This suggests that regardless of global environment, galaxies living in dense local environments tend to be more massive.
We also note that this aligns approximately with the local densities where we see the greatest change to morphological fractions in Figure \ref{fig:MDR_cl+fd_kernel},
confirming that galaxy morphology, local density, and stellar mass are all strongly correlated.

However, in the right panel of Figure \ref{fig:density-mass}, we see that the cluster and field samples have different distributions of local density with galaxy stellar mass.
While the median local density of cluster galaxies increases with stellar mass (from $\sim12$ to $36$ gal Mpc$^{-2}$), the median local density of field galaxies is invariant with stellar mass, remaining constant at $\sim2$ gal Mpc$^{-2}$.
Comparing these values against Figure \ref{fig:MDR_cl+fd_kernel} and Figure \ref{fig:median_sersics}, we do see differences in both the morphological fractions and median Sérsic index of galaxies at these densities.
Therefore, if significant differences in the morphology-mass relation between cluster and field samples do exist, it is plausible that they may be explained by differences in their typical local densities.

With this in mind, we can infer from the results of Figure \ref{fig:density-mass} that the morphology-mass relation in the field sample is primarily representative self-evolution, i.e. the evolution of galaxy morphology with stellar mass in the absence of environmental influence. If the morphology-mass relation in the cluster sample is indeed in disagreement with this trend, it may be indicative of \textit{local} environmental influence beyond what is expected from morphological evolution with stellar mass alone.


\section{Discussion}
\label{sec:discussion}


\subsection{Does Environment or Stellar Mass Drive Morphological Evolution at Redshift 1.6?}
\label{subsec:physical_origins}

We have observed a morphology-density relation at $z\sim1.6$, such that the fraction of bulge-like galaxies increases with local galaxy density, while the fraction of disk-like galaxies decreases. This is supported by our finding of a more general structure-density relation, such that the median Sérsic index of galaxies increases smoothly with local density. We observe these relations in both our cluster and field samples with good similarity. Regardless of global environment, the majority of galaxies at low density are disk-like ($n<1.5$), while at high density, an equal proportion of galaxies are bulge-like ($n>2.5$). Our findings provide new insights into what drives of the morphology-density relation at this epoch.

Many studies have favored cluster-specific processes as the main driver of the morphology-density relation. Environmental processes such as ram pressure stripping \citep{Gunn1972} and strangulation \citep{Larson1980,Balogh2000} can quench galaxies as they move through the hot intracluster medium or fall into a cluster's gravitational well, while galaxy harassment \citep{Moore1996} can disturb morphologies through repeated, high-speed flyby interactions. However, the morphology-density and structure-density relations in our cluster sample show strong similarity to those of the field. This suggests that the same physical process(es) may drive the relations in both samples, which rules out processes specific to the cluster environment.

Given the correlation between galaxy Sérsic index and stellar mass \citep{vanderWel2008, D'Onofrio2015}, it is reasonable to question what role stellar mass has to play in the morphology-density relation.
As shown in Section \ref{subsec:morph-mass}, we find that galaxy morphology and structure are indeed correlated with increasing stellar mass, as they are with local galaxy density.
Our results show that high stellar masses ($M_* \gtrsim 10^{10.8} M_{\odot}$) are required to build up a significant fraction of bulge-like ($n>2.5$) galaxies, regardless of local density or global environment.
Additionally, in both cluster and field samples, the growth in median stellar mass and the convergence of morphological fractions appear to occur at similar values of local galaxy density.
However, as shown in Section \ref{subsec:morph-dens}, there is no significant difference in the morphology-density and structure-density relations between our cluster and field samples, even though the samples have different distributions of galaxy stellar masses.
In the nearby universe, \citet{Fasano2015} find that the relationships of morphology with local density and clustercentric radius are both strong even in narrow ranges of galaxy stellar mass.
While at higher redshift, our results seem to agree that perhaps stellar mass is not the only driver in the observed morphology-density relation at this redshift.

We do, however, find there may be evidence of differences between cluster and field when it comes to the morphology-mass and structure-mass relations.
The slopes of the morphology-mass relation appear steeper in the cluster sample, and cluster galaxies have higher median Sérsic indices than field galaxies at both intermediate and high stellar masses ($M_* \gtrsim 10^{10.25} M_\odot$).
The majority of our statistical tests suggest there may be less than a 50 percent chance that the morphology-mass relations are consistent between cluster and field samples.
If these relations are truly different, they may be impacted by environmental differences.
Local density is positively correlated with stellar mass in the cluster sample, indicating that massive galaxies tend to reside in the densest regions of the cluster.
In contrast, field galaxies show no correlation of local density with stellar mass.
However, these interpretations should be taken lightly, as we cannot be confident that the observed differences in the morphology-mass relation are significant.
Larger samples of galaxies, particularly in clusters, would be required to further investigate these trends, as well as the separability of mass and environment with morphology.
For now, the exact role of stellar mass remains unclear.

\subsection{The Physical Origins of the Morphology-Density Relation}

The presence of a morphology-density relation as early as $z\sim1.6$ suggests that the morphology-density relation may be baked into galaxy formation at early times.
In a study of low redshift galaxies from SDSS, \citet{Wang2018} find that halo mass is the primary predictor for the fraction of quenched galaxies in groups. They show that a model using a combination of host halo mass, galaxy stellar mass, and halo assembly bias can fully account for the observed trends in quenched fraction with environmental mass density in their sample.
Using cosmological simulations to study galaxies from $1.5\leq z \leq3.5$, \citet{Ahad2024} find that properties of galaxy halos such as peak circular velocity and total halo mass are stronger indicators of quenching than galaxy stellar mass alone, regardless of cluster or field environment. By making comparisons between central and satellite galaxies, both studies suggest that what has previously been considered an environmental dependence on quenching may actually be attributed to halo properties.

This is in agreement with observational results from \citet{Nantais2017}, who suggest that quenched fractions in cluster and field galaxies at $z\sim1.6$ may be influenced by both total halo mass and halo age.
Many studies have found good correlation between galaxy quiescence and Sérsic index up to high redshifts \citep{Bell2012, Kawinwanichakij2017, Matharu2019}.
If morphology follows the same trends as quiescence, it may be possible that galaxy morphology is pre-determined at early times by the dark matter halo in which the galaxy resides.
This may also explain why galaxies in $z\sim1.6$ \textit{field} overdensities appear to be structurally consistent with those in \textit{cluster} regions, as their dark matter halos may already have similar properties to protocluster environments.

Group pre-processing is a scenario in which galaxy properties, such as morphology, are first transformed in a group environment which then falls into the cluster at later times \citep{Wetzel2013}.
Pre-processing may be accomplished through an increase in galaxy mergers.
Although major mergers are considered to be one of the primary mechanisms responsible for the formation of elliptical galaxies \citep{Mihos1994}, they are expected to be less common in clusters due to high velocity dispersions \citep{Just2010, Bekki2011, Mei2012}.
Indeed, \citet{Delahaye2017} find that the merger rates in our $z\sim1.6$ clusters are roughly the same or lower than that of the field.
Galaxy groups provide an environment with higher densities than the field and lower velocity dispersions than clusters, making them ideal for mergers.
\citet{Sazonova2020} also find that group pre-processing is a possible dominant mechanism driving the morphology-density relation at $z\sim1.5$.
Alternatively, \citet{Delahaye2017} propose that protoclusters may also be a favorable environment for merging and pre-processing galaxies into bulge-dominated morphologies.
If the morphology-density relation is driven primarily by pre-processing at $z\sim1.6$, it could explain the observed similarity between field overdensities and our clusters.

At low redshift, studies such as \citet{Goto2003} and \citet{Vulcani2023} have found evidence of multiple mechanisms driving the overall shape of the morphology-density relation. 
While \citet{Goto2003} suggest two different mechanisms that act on different environmental scales (one of which depends on the cluster environment), \citet{Vulcani2023} suggest that most elliptical galaxies trace a ``primordial'' density distribution, while the ratio of spirals to lenticulars show evidence of morphological transformation by the cluster environment.
With this in mind, the fact that our $z\sim1.6$ clusters display such a similar morphology-density relation to the field can be attributed to two main possibilities.

Firstly, our clusters may be less evolved than low-redshift counterparts.
\citet{Nantais2017} find that quenched fractions in J0330, J0225, and J0224 are only marginally higher than the field at $z\sim1.6$, implying a significantly lower quenching efficiency than observed in clusters at $z\lesssim1.3$.
While \citet{Noble2019} and \citet{Cramer2023} find evidence of one-sided gas tails in small samples of cluster galaxies, an analysis of galaxy color gradients by \citet{Cramer2024} finds no strong evidence of widespread quenching by ram pressure stripping.
\citet{Delahaye2017} and \citet{Nantais2020} also both suggest that these clusters are likely not yet virialized.

Secondly, the dominant physical processes driving the morphology-density relation in the high-redshift universe may be different than those in the low-redshift universe.
Previous work has found evidence of an evolution in the relation between $z\sim1$ and present day \citep{Postman2005, Smith2005, Tasca2009, Euclid2025}, which could suggest a widespread change in the mechanisms responsible for driving the relation over cosmic time.
If we consider that many protoclusters only just beginning to virialize during cosmic noon, it follows that environmental mechanisms relying on dynamically-evolved states or long timescales may be less efficient at transforming galaxy morphology during this epoch.
Therefore, it is reasonable to expect that the morphology of cluster galaxies may be more similar to that of field galaxies at this epoch and earlier cosmic times.

Since no other studies have performed an equivalent comparison between cluster and field galaxies, it is difficult to know how our findings fit into the broader picture of galaxy evolution.
While we do not believe our findings to be specific to the three clusters sampled in this work, it would be valuable to analyze a greater number of galaxy clusters at the epoch of cosmic noon to confirm whether these trends continue to be observed across a wider range of large-scale environments.
Another important question that remains is whether the similarity we have observed between the morphology-density relations of cluster and field galaxies disappears at lower redshift, and what this could tell us about the mechanisms that drive environmental galaxy evolution across cosmic time.
Determining the evolution of the morphology-density relation will take a wide range of cluster and field samples across a range of redshifts, along with self-consistent methods and data analysis.
We aim to address these questions in future work.

\subsection{Comparison with Other Studies}
\label{subsec:other_studies}

Our findings on the morphology-density relation at $z\sim1.6$ are in good agreement with other studies of galaxy clusters at this epoch.
\citet{Mei2023} find that the fraction of visually-identified early-type galaxies increases with local density in clusters and protoclusters from $1.4<z<2.8$.
While they do not measure local galaxy density, \citet{Sazonova2020} find that their $z=1.75$ cluster not only has a higher fraction of bulge-dominated galaxies than the field, but also that galaxies in the inner region of the cluster tend to have higher Sérsic indices than galaxies in the outer region (although the latter result does not hold true for their $z=1.19$ cluster).
From $1.4<z<1.7$, \citet{Strazzullo2023} also find that the central regions of clusters host a higher fraction of bulge-dominated ($n>2.5$) galaxies than the field.
We note that we are the only study at this epoch to calculate projected local density from galaxy redshifts, as well as the only study to separately measure and compare the morphology-density relation in both cluster and field samples.

We also find general agreement with other studies on the relationship between galaxy morphology and stellar mass at this epoch.
As mentioned, \citet{Strazzullo2023} find that cluster galaxies above their mass limit of $M_*>10^{10.85} M_{\odot}$ are predominantly bulge-dominated, while field galaxies of the same mass are more evenly distributed between bulge- and disk-dominated ($n<1.5$).
In a $z\sim1.6$ cluster in UDS, \citet{Bassett2013} find that $n>2$ galaxies are more likely to be high-mass and quenched, although they do not comment directly on differences with environment.
\citet{Mei2023} find a trend of increasing early-type galaxy fraction with stellar mass in their clusters, however, this result is found with only 2 stellar mass bins.
While \citet{Sazonova2020} do not find evidence that Sérsic index increases with stellar mass in their $z=1.75$ cluster, they do find that low-mass ($M_*\lesssim 10^{10.5} M_{\odot}$) cluster galaxies have higher Sérsic indices than their field counterparts.
Both \citet{Mei2023} and \citet{Strazzullo2023} caution that their stellar mass estimates have large uncertainties due to limited photometry and IR coverage of their clusters, which further highlights the importance of the dataset used in our work.

For greater context, we also compare our findings on the morphology-mass relation to those from lower-redshift studies.
From $0.2<z<1$, \citet{Tasca2009} find that the morphology-density relation is present for galaxies with $M_*>10^{10.6} M_{\odot}$, but that above this mass, morphology is independent of local density, suggesting that environmental influences on morphology may be dominant at lower stellar masses.
\citet{Euclid2025} find that the fraction of early-type galaxies increases with stellar mass across all redshift bins from $0.25<z<1$.
In their highest redshift bin ($0.75<z<1$), the morphology-density relation is strongest for low-mass ($M_*<10^{10.5} M_{\odot}$) galaxies, while high-mass galaxies are already predominantly early-type regardless of local density.
\citet{Vulcani2011} find that for both $0.04 \leq z \leq 0.07$ and $0.4 \leq z \leq 0.8$ clusters, elliptical galaxies are the dominant morphology above $10^{11} M_{\odot}$.
Across their stellar mass range, they find that the fraction of elliptical galaxies is similar for both redshifts, and relatively flat for lower stellar masses.
In almost all environments at $z\lesssim0.1$, \citet{Calvi2012} find an increase in elliptical galaxies and a decrease in S0 and late-type galaxies with $M_*>10^{11} M_{\odot}$.
However, for $M_*<10^{11} M_{\odot}$, non-cluster environments show only a weak dependence on stellar mass.
Although at higher redshift, our results are generally in agreement with these findings.


\section{Summary and Conclusions}
\label{sec:conclusions}


We study the morphology-density, morphology-mass, and density-mass relationships using a sample of galaxies from three SpARCS clusters and two 3D-HST/CANDELS fields at $z\sim1.6$.

We select galaxies with $\log(M_*/M_{\odot}) \geq 9.58$ and fit their F160W (rest-frame R-band) light profiles to single-component Sérsic profiles using \textsc{Galfit}. Using Sérsic index ($n$) as a proxy for morphology, we define galaxies with $n<1.5$ as disk-like, $1.5 \leq n \leq 2.5$ as intermediate, and $n>2.5$ as bulge-like. We calculate projected local galaxy densities using 5th-Nearest Neighbors within a line-of-sight velocity slice of $\Delta v=\pm3000$ km s$^{-1}$, using a combination of spectroscopic, grism, and photometric galaxy redshifts.

Through comparison of cluster and field galaxies, we find the following at $z\sim1.6$:

\begin{enumerate}
    \item A morphology-density relation is already in-place. (Figure \ref{fig:MDR_cl+fd_binned} and \ref{fig:MDR_cl+fd_kernel})
    \item The morphology-density and structure-density relations are similar in both cluster and field galaxies, suggesting these relations may be independent of global environment (i.e. not specific to clusters) at this epoch. (Figure \ref{fig:median_sersics}, \ref{fig:MDR_unified_binned}, and \ref{fig:MDR_unified_kernel})
    \item In general, both cluster and field galaxies show a positive correlation between Sérsic index and galaxy stellar mass. High stellar masses ($M_* \gtrsim 10^{10.8} M_{\odot}$) are required to build up a significant fraction of bulge-like galaxies, regardless of environment. (Figure \ref{fig:median_sersic-mass})
    \item Cluster and field galaxies may follow slightly different morphology-mass and structure-mass relations, such that stellar mass may have a stronger impact on the morphology of low- and intermediate-mass cluster galaxies compared to field galaxies of the same mass. Disk-like galaxies remain the majority until $M_* \gtrsim 10^{10.8} M_{\odot}$ in the field, but only $M_* \gtrsim 10^{10.25} M_{\odot}$ in the cluster. (Figure \ref{fig:MMR_kernel} and \ref{fig:median_sersic-mass})
    \item While stellar mass is similarly correlated with local galaxy density in both cluster and field galaxies, local density is only positively correlated with stellar mass in clusters.
    High-mass cluster galaxies preferentially reside in denser regions of the cluster, while high-mass field galaxies show no preference in density compared to low-mass field galaxies. (Figure \ref{fig:density-mass})
\end{enumerate}

The discovery of an unambiguous morphology-density relation at $z\sim1.6$ begs the question of when in cosmic time the relation is first established.
New high-redshift datasets such as those from JWST will provide the capability to answer this question, and perhaps fundamentally transform our understanding of the relationship between galaxy morphology, mass, and environment.


\begin{acknowledgments}
This research is based on observations made with the NASA/ESA Hubble Space Telescope obtained from the Space Telescope Science Institute, which is operated by the Association of Universities for Research in Astronomy, Inc., under NASA contract NAS 5–26555. These observations are associated with programs GO-13306 and GO-13677/14327.
This work is based in part on observations taken by the 3D-HST Treasury Program (GO 12177 and 12328) with the NASA/ESA HST, which is operated by the Association of Universities for Research in Astronomy, Inc., under NASA contract NAS5-26555. 
G.W. gratefully acknowledges support from the National Science Foundation through grant AST-2347348.
R.D. gratefully acknowledges support by the ANID BASAL project FB210003.
\end{acknowledgments}



\software{Source Extractor \citep{Bertin1996},
          Grizli \citep{Grizli},
          GALFIT \citep{Peng2002,Peng2010a}
          }


\bibliography{references}{}
\bibliographystyle{aasjournal}


\end{document}